\documentclass{article}
\usepackage{amsfonts,amsmath,amssymb}
\usepackage{mathtools}
\usepackage{bbm}
\usepackage{dsfont}
\usepackage{multirow}
\usepackage{setspace}
\usepackage{algpseudocode}
\usepackage{algorithm,algorithmicx}
\usepackage{color}
\usepackage[pdftex]{graphicx}
\usepackage{mathptmx}
\usepackage{comment}
\usepackage{float}
\usepackage[font = small, labelfont=bf, labelsep=space]{caption}
\usepackage{subcaption}
\usepackage{here}
\usepackage[authoryear]{natbib}
\usepackage[colorlinks = true, citecolor = orange]{hyperref}
\usepackage{tikz}
\usepackage{standalone}
\usetikzlibrary{decorations.pathmorphing}
\usetikzlibrary{decorations.pathreplacing}
\usepackage{place ins} 
\newlength{\defbaselineskip}
\DeclareMathOperator{\Pb}{\mathbb{P}}

\newcommand{\ds}{\displaystyle}

\definecolor{greenish}{rgb}{.2, .8, .4}

\allowdisplaybreaks

\makeindex

\makeatletter

\title{An efficient distribution method for nonlinear two-phase flow in highly heterogeneous multidimensional stochastic porous media}

\author{Fayadhoi Ibrahima 
  \thanks{Institute for Computational \& Mathematical Engineering, Stanford University, Stanford, CA}
\and Hamdi A. Tchelepi 
  \thanks{Department of Energy Resources Engineering, Stanford University, Stanford, CA, USA}
\and Daniel W. Meyer
  \thanks{ Department of Mechanical and Process Engineering, Institute of Fluid Dynamics, ETH Z\"urich, Z\"urich, Switzerland}
}
\begin{document}

\maketitle

\begin{abstract}

	In the context of stochastic two-phase flow in porous media, we introduce a novel and efficient method to estimate the probability distribution of the wetting saturation field under uncertain rock properties in highly heterogeneous porous systems, where streamline patterns are dominated by permeability heterogeneity, and for slow displacement processes (viscosity ratio close to unity). Our method, referred to as the frozen streamline distribution method (FROST), is based on a physical understanding of the stochastic problem. Indeed, we identify key random fields that guide the wetting saturation variability, namely fluid particle times of flight and injection times. By comparing saturation statistics against full-physics Monte Carlo simulations, we illustrate how this simple, yet accurate FROST method performs under the preliminary approximation of frozen streamlines. Further, we inspect the performance of an accelerated FROST variant that relies on a simplification about injection time statistics. Finally, we introduce how quantiles of saturation can be efficiently computed within the FROST framework, hence leading to robust uncertainty assessment.

\end{abstract}

\section{Introduction}

The need for assessing and quantifying uncertainty in subsurface flow has driven research in the stochastic aspects of hydrology and multiphase flow physics. Regardless of whether the interest lies in simulating the spreading of a contaminant in aquifers or predicting the oil recovery in a reservoir, precise and robust numerical methods have been developed to provide quantitative answers for these flow responses. However, the complex nature of subsurface systems, together with inherent incomplete information about their properties, have resulted in the surge of probabilistic modeling of these uncertainties. The uncertainty propagation from input subsurface properties to output flows is then naturally recast into a stochastic partial differential equation (PDE) problem. Within this stochastic context, significant work have addressed the estimation of the response flow statistics. The fundamental work of \cite{G}, \cite{Dag} and \cite{Z} have demonstrated how geostatistical treatments of subsurface properties can be used to derive statistical information of the flow response. These statistics (generally mean and variance) have been estimated in different ways. 

Statistical Moment Equation (SME) methods, also known as Low Order Approximations (LOA), build deterministic PDEs for the moments by averaging stochastic PDEs. \cite{LT} used these methods to estimate the statistical moments of the pressure field for single phase flow under uncertain permeability field. However, since this class of methods relies on perturbations, the moment estimates are only satisfactory for very small variances, which limits their applicability towards uncertainty quantification.
Stochastic spectral methods are another class of popular methods to estimate statistical moments \citep[see e.g.][]{BNT, BNTT} and rely on the truncation of the Karhunen-Lo\`eve (KL) expansion of random processes \citep{L}. Stochastic spectral methods can lead to rigorous convergence analysis \citep[e.g., ][]{C}, but they suffer from an exponentially increasing complexity with the dimension and the correlation scales of the problem (this is known as the curse of dimensionality). Furthermore, they require further sophistications for highly nonlinear problems \citep[see e.g.][]{A,FWK,LKNG}.
Finally, distribution methods aim at estimating the Probability Density Function (PDF) and Cumulative Distribution Function (CDF) of the flow response usually by solving deterministic PDEs for the distributions. These methods are attractive because they offer a complete picture of the variability of the flow response. Distribution methods have led to an abundant literature, especially for the study of stochastic turbulent flows \citep[e.g.,][]{P} or more recently for tracer flows \citep[e.g.,][]{MJT, MT, MTJ} and advection-reaction problems \citep[e.g.,][]{TB,VTTK}. However, these methods typically rely on closure approximations and hence may not be generally applicable.   
The most popular and reliable method in uncertainty propagation remains Monte Carlo Simulation (MCS). However, due to its slow convergence rate, it requires a prohibitively large number of samples to be accurate for large-scale applications, such as reservoir simulation and hydrology \citep{BG}. 

For stochastic two-phase flow in heterogeneous porous media, which is what we are discussing in this paper with the Buckley-Leverett (BL) model \citep{BL, Peace, AS}, additional challenges occur. Indeed, the governing transport equation is nonlinear due to the coupling between the saturation and the total velocity, as well as the presence of fractional flows. This nonlinearity affects the quality of the stochastic methods aforementioned. For the stochastic BL problem, to provide first and second moments of the water saturation, \cite{ZT} and \cite{ZLT} had to assume that the distribution of the travel time was known, while still relying on an LOA approach. The key of their approach, though, was to identify the travel time as the underlying random field that explains the saturation statistics, and hence give a physical insight of the stochastic flow problem. Instead, \cite{JR} used an Eulerian formulation to express first and second order moment SMEs for the BL problem. They showed that the solution develops bimodal behaviors that are non-physical, thus demonstrating the potential fragility of SME methods for nonlinear flow problems when no physical information is exploited (see also \cite{LLZ} and \cite{JT}). For spectral methods, \cite{LZ1} and \cite{LZ2} used physically motivated transforms to build efficient probabilistic collocation methods for nonlinear flow in porous media. For MCS, \cite{MJM} show that Multilevel Monte Carlo (MLMC) methods, together with the use of streamlines, speed up the naive MCS method when estimating the moments of the water saturation under uncertain lognormal permeability fields. Regarding distribution-based strategies, \cite{WTJT} extended the ideas of \cite{TB} for the advective-reactive transport problem to successfully build a CDF framework for the one-dimensional BL case with known distribution for the total flux. However there is no clear extension to multiple spatial dimensions and little physical insight is explored. 
Furthermore, if often the permeability field is taken to be log normally distributed in the literature, (which, according to \cite{GO} and \cite{COZ}, does not prevent the saturation distribution to be highly non-Gaussian,) a geostatistical modeling conveys more realistic scenarios \citep[see e.g.][]{Deutsch, DJ}. This alternative stochastic modeling, however, reinforces the pitfalls of the aforementioned methods since no Gaussian assumption is valid for the closure-based SME methods, and the covariance function used for KL expansions in stochastic spectral methods can require a prohibitively large number of KL terms to be informative.   

Because distribution methods provide a complete description of the stochastic variability of random fields via the PDF and CDF estimates, and because this full description is important to describe the complexity of the stochastic saturation field, our intent is to built efficient distribution methods for the saturation, especially for high variance scenarios. In addition, we are interested in exploiting the physics of the problem to reach our goal.  

We propose a new distribution method, called the FROST, to treat the stochastic two-phase flow problem. The FROST is an extension of the work by \cite{IMT} to higher spatial dimensions, where a similar approach was first derived for one-dimensional two-phase flow. The FROST produces reliable and fast estimates of the PDF and CDF of the (water) saturation by using a flow-driven approach to express the saturation as a (semi-analytical) nonlinear function of a smooth random field. The FROST offers saturation distribution estimates that are suitable for high permeability variances, and that stay reliable with geostatistically driven permeability fields. What's more, the saturation PDF estimates, though discontinuous, stay stable by construction, while direct Kernel Density Estimation (KDE) with MCS fails to converge or converge at very slow rates \citep{CH, WW}. Finally, since the saturation field has a complex distribution, the study of probability of exceedance \citep{DN} rather than first statistical moments appears to be more suited for uncertainty quantification. The FROST also leads to fast computable probabilities of exceedance, which we refer to as saturation quantiles. \\

The paper proceeds as follows. First, \autoref{section : reduc1D} introduces the two-phase flow problem and recalls the main sources of uncertainties considered in this study. Then, \autoref{section : methodology} describes how, as a critical first step, we use the flow velocity to cast the transport problem into multiple one-dimensional problems along streamlines, following the methodology adopted in \cite{IMT} to derive saturation PDF and CDF for one-dimensional problems. We then take an extra step and extend the methodology to be applicable to two or three dimensional porous media. In these multi-dimensional stochastic porous media, we show that the saturation PDF/CDF can be estimated based on the statistics of two random fields, the time of flight (TOF) --- the travel time of a particle to a given position, and the equivalent injection time (EIT) --- an injection time accounting for changes in streamtube capacity, that can both be efficiently estimated. Moreover, the FROST relies on the approximation of frozen streamlines over time, which is applicable for highly heterogeneous porous media. This approximation leads to a unique and deterministic mapping between the EIT and the TOF on the one hand, and the saturation on the other hand. Consequently, \autoref{section : companal} showcases the accuracy and robustness of the FROST approach. More specifically, we assess the different approximations made in our method by means of numerical tests and sensitivities against MCS. In addition, the analysis leads to the proposal of a faster FROST implementation, named gFROST, where the TOF distribution is approximated by a log-Gaussian random field. To further illustrate the performance of the FROST, \autoref{section : numres} gathers numerical comparisons of the average and standard deviation of the saturation for more challenging stochastic permeability fields. In addition, we provide the corresponding saturation quantiles from the FROST. Finally, this paper closes with a discussion on the method and results in 
\autoref{section : disc}, along with some concluding remarks in \autoref{section : conclusion}.

\section{Problem formulation}  
\label{section : reduc1D}

To simulate two-phase flow in porous media, \cite{AS} demonstrated that the BL equation proves to be a simple but 
practical enough model. We assume that we have a wetting phase (e.g. water) and a non-wetting phase (e.g. oil), and we are interested in estimating the water saturation field in the porous medium, i.e. the fraction of water in a given pore space. The BL equation uses a few parameters to model the saturation transports in porous media. These parameters are the porosity $\phi(\vec{x})$ and the permeability field 
$K(\vec{x})$; the water and oil viscosities $\mu_w$ and $\mu_o$, and the pressure (resp. rate) on controlled zones (injectors, producers) $p_{inj}$ (resp. $\mathbf{q}_{inj}$) and $p_{prod}$ (resp. $\mathbf{q}_{inj}$); the relative permeability fields 
$k_{rw}(S_w)$ and $k_{ro}(S_o)$ for the interaction between the fluids and the medium. We assume no chemical reactions and no change of state. 

In the ideal case, the prediction of the saturation field relies on a complete knowledge of the aforementioned parameters. While experimental estimations of fluid parameters are accessible, due to limited measurements in the subsurface, a lack of information on the rock properties is usually the dominant cause of uncertainty in the flow response. We will therefore focus on uncertainties in the porosity and permeability fields in the present paper. We model these uncertainties by assigning probabilistic distributions to these fields that reflect many likely scenarios. The goal of our study is to estimate the probabilistic distribution of the saturation field under these stochastic porosity and permeability fields.
The saturation field is governed by a coupled system of equations: a nonlinear transport equation (mass conservation) and a Darcy equation (pressure equation). The total Darcy flux or velocity $\mathbf{q}_{tot}$ is defined as the sum of Darcy fluxes for each phase, and relates the discharge rate through 
the porous medium to the gradient of pressure in a proportional manner involving the previously defined parameters, i.e.
\begin{eqnarray}
\label{eq:Darcy}
	\mathbf{q}_{tot} & = & \mathbf{q}_{w} + \mathbf{q}_{o} \mbox{ with} \nonumber \\ 
	\mathbf{q}_{w} & = & -K(\vec{x}) \frac{k_{rw}(S_w)}{\mu_w} \nabla p = -\lambda_{w} \nabla p \mbox{ and} \\
	\mathbf{q}_{o} & = & -K(\vec{x}) \frac{k_{ro}(S_o)}{\mu_o} \nabla p = -\lambda_{o} \nabla p. \nonumber
\end{eqnarray} 
The total Darcy flux is assumed to verify the incompressibility condition, 
\begin{equation}
\label{eq:incompressible}
	\nabla \cdot \mathbf{q}_{tot}  = 0.
\end{equation} 
Furthermore, we can define the water fractional flow $f_w = \lambda_w / (\lambda_w + \lambda_o)$ and the viscosity ratio $m = \mu_w/\mu_o$, so that 
\begin{displaymath}
	\mathbf{q}_{w}  = f_w  \mathbf{q}_{tot} \mbox{ and } f_w = \frac{k_{rw}}{k_{rw} + m k_{ro}}.
\end{displaymath} 
With the condition $S_w + S_o = 1$, the conservation of mass can be reduced to one equation solely involving the water saturation,  
\begin{equation}
\label{eq:consmass}
	\phi(\vec{x}) \frac{\partial S_w}{\partial t} +\mathbf{q}_{tot} \cdot \nabla f_w(S_w)  = 0.
\end{equation} 
Finally, we have boundary conditions either on the pressure field (pressure control),  
\begin{equation}
\label{eq:BCpressure}
\begin{array}{rl}
	p(\vec{x}) = p_{inj}, & \vec{x} \in \Gamma_{\mbox{\small{inj}}}, \\
	p(\vec{x}) = p_{prod}, & \vec{x} \in \Gamma_{\mbox{\small{prod}}}, \\
\end{array}
\end{equation}
where $p_{inj}$ and $p_{prod}$ are respectively prescribed pressures at the injector and at the producer, 
or on the injection/pumping rates (rate control), 
\begin{displaymath}
\begin{array}{rl}
	\mathbf{q}_{tot}(\vec{x}) = \mathbf{q}_{inj}, & \vec{x} \in \Gamma_{\mbox{\small{inj}}}, \\
	\mathbf{q}_{tot}(\vec{x}) = \mathbf{q}_{prod}, & \vec{x} \in \Gamma_{\mbox{\small{prod}}}, \\
\end{array}
\end{displaymath}
as well as initial/boundary conditions on the water saturation, 
\begin{displaymath}
\begin{array}{rl}
	S_w(\vec{x}, t) = 1 - s_{oi}, & \vec{x} \in \Gamma_{\mbox{\small{inj}}} \mbox{ and } t > 0, \\
	S_w(\vec{x}, 0) = s_{wi}, & \vec{x} \in \Omega, \\
\end{array}
\end{displaymath}
where $\Omega$ is the reservoir, $\Gamma_{\mbox{\small{inj}}}$ is the boundary of the reservoir where the water is injected from the 
injector, $\Gamma_{\mbox{\small{prod}}}$ is the boundary of the reservoir where water and oil are collected from the producer, and $s_{wi}$ and $s_{oi}$ are respectively the irreducible saturations of water and oil. 

\medskip

\section{Methodology}
\label{section : methodology}

\subsection{Saturation distribution estimation via stochastic streamlines}
\label{subsection:streamlines}

\begin{figure}
	\centering
	\begin{tikzpicture}[scale=4,>=to,line width=1.5pt,every text node part/.style={align=center}]

\draw [->](0,0) -- (0,1); 
\draw [->](0,0) -- (1,0); 
\draw [-](0,1) -- (1,1); 
\draw [-](1,0) -- (1,1); 

\draw (1,-0.1) node(text1){$\ds \mathbf{x}_{1}$};
\draw (-0.1,1) node(text1){$\ds \mathbf{x}_{2}$};

\draw (0,-0.1) node(text1){$\ds \mathbf{x}_{sl}(0)$};
\draw[gray] (-0.12,0.1) node(text2){$0$};
\draw[blue] (-0.2,0.0) node(txt2){(Injection)};
\draw[gray] (0.88,1.1) node(text3){$L$};
\draw[green] (1.2,1.1) node(txt3){(Production)};

\draw[gray] (0.38,.8) node(text4){$r(\tau(\vec{x}))$};
\draw[red] (0.68,0.64) node(text5){$\mathbf{r} = \mathbf{x}_{sl}(\tau)$};

\draw[gray] (0,0) -- (-0.05,0.05);
\draw[gray] (1,1) -- (0.95,1.05); 	
\draw[gray] (.55,.7) -- (0.5,.75); 


\draw[red] (0.1,1.1) node(txt5){streamlines};
\draw[red, thin] (txt5) -- (0.26,.88);
\draw[red, thin] (txt5) -- (0.2,.5);
\draw[red, thin] (txt5) -- (0.4,.16);

\draw[red, very thick] (0, 0) to [out=70, in=160] (.4,.85); 
\draw[->,red, very thick] (.4, .85) to [out=-20, in=220] (.5,.9); 
\draw[red, very thick] (.5, .9) to [out=200, in=-178] (.7,.95); 
\draw[red, very thick] (.7, .95) to [out=-5, in=180] (1,1); 

\draw[red, thick] (0, 0) to [out=30, in=160] (.3,.5); 
\draw[->,red, thick] (.3, .5) to [out=-20, in=-178] (.55,.7); 
\draw[red, thick] (.55, .7) to [out=2, in=3] (.6,.8); 
\draw[red, thick] (.6, .8) to [out=177, in=-178] (.8,.85); 
\draw[red, thick] (.8, .85) to [out=-5, in=190] (1,1); 

\draw[gray, thick] (-.05, .05) to [out=30, in=160] (.25,.55); 
\draw[->,gray, thick] (.25, .55) to [out=-20, in=-178] (.5,.75); 
\draw[gray, dotted] (.5, .75) to [out=2, in=3] (.55,.85); 
\draw[gray, dotted] (.55, .85) to [out=177, in=-178] (.75,.9); 
\draw[gray, dotted] (.75, .9) to [out=-5, in=190] (.95,1.05); 

\draw[gray!80, thick, nearly transparent] (0, 0) to [out=10, in=160] (.65,.2); 
\draw[gray!80, thick, nearly transparent] (.65, .2) to [out=-20, in=-178] (.8,.23); 
\draw[gray!80, thick, nearly transparent] (.8, .23) to [out=2, in=3] (.75,.3); 
\draw[gray!80, thick, nearly transparent] (.75, .3) to [out=177, in=-178] (.82,.5); 
\draw[gray!80, thick, nearly transparent] (.82, .5) to [out=-5, in=190] (1,1); 

\draw[->,red, very thick] (0, 0) to [out=10, in=160] (0.675,.15); 
\draw[red, very thick] (.675, .15) to [out=-20, in=-178] (.85,.215); 
\draw[red, very thick] (.85, .215) to [out=2, in=3] (.815,.3); 
\draw[red, very thick] (.815, .3) to [out=177, in=-178] (.87,.475); 
\draw[red, very thick] (.87, .475) to [out=-5, in=190] (1,1); 

\draw[gray!80, thick, nearly transparent] (0, 0) to [out=10, in=160] (.7,.1); 
\draw[gray!80, thick, nearly transparent] (.7, .1) to [out=-20, in=-178] (.9,.2); 
\draw[gray!80, thick, nearly transparent] (.9, .2) to [out=2, in=3] (.88,.3); 
\draw[gray!80, thick, nearly transparent] (.88, .3) to [out=177, in=-178] (.92,.45); 
\draw[gray!80, thick, nearly transparent] (.92, .45) to [out=-5, in=190] (1,1); 

\begin{scope}
\clip (0, 0)  to [out=10, in=160] (.65,.2) 
	 to [out=-20, in=-178] (.8,.23) 
	 to [out=2, in=3] (.75,.3) 
	 to [out=177, in=-178] (.82,.5) 
	 to [out=-5, in=190] (1,1)
	 -- (1,0)
	 -- (0,0); 
\clip (0, 0) to [out=10, in=160] (.7,.1) 
	 to [out=-20, in=-178] (.9,.2) 
	 to [out=2, in=3] (.88,.3)
	 to [out=177, in=-178] (.92,.45) 
	  to [out=-5, in=190] (1,1)
	 -- (0,1)
	 -- (0,0); 
	\draw[fill=gray!80, nearly transparent]	(0,0) rectangle (1,1); 
\end{scope}

\draw[black, dashed] (0.75,0.45) -- (0.86,0.38); 
\draw[black] (0.68,.45) node(txt6){$A$};

\draw[gray!80] (1.25,0.1) node(txt8){streamtube};
\draw[gray!80, thin] (txt8) -- (0.9,.2);

\end{tikzpicture}
	\caption{Illustration of streamline related notations. For each streamline, $\vec{x}_{sl}(\tau)$ corresponds to the position of a tracer at travel time $\tau$ and $r(\tau)$ is its travel distance, which goes from $0$ at the injection to $L$, the final length at the production end. Also, for each streamtube, $A(r)$ corresponds to its cross-section area at distance $r$.}
	\label{fig:streamline}
\end{figure}
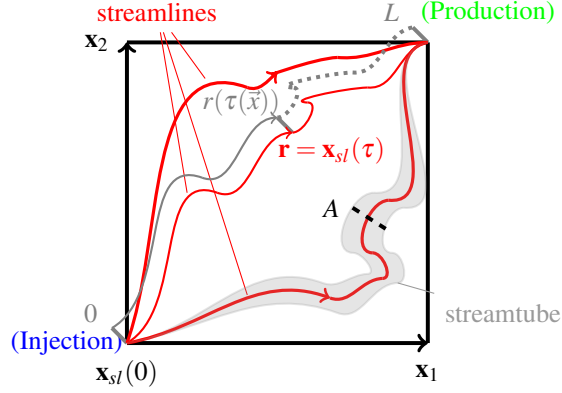
We can recast \autoref{eq:consmass} into a 1D framework by considering the following characteristic curves or streamlines \citep[e.g.,][]{MW, CD, TBBO},  
\begin{equation}
\label{eq:cons}
	\vec{x}_{sl}(\tau) = \vec{x}_{0, sl} + \int_0^{\tau} \mathbf{q}_{tot}[\vec{x}_{sl}(t'), t'] dt'
\end{equation}
with $\tau$ being the TOF and $\vec{x}_{0, sl}$ being the origin of the streamline at $\tau = 0$. Considering the directional derivative along a streamline, we can write
\begin{displaymath}
	\mathbf{q}_{tot} \cdot \nabla = \left| \mathbf{q}_{tot}\right| \frac{\partial}{\partial r}
\end{displaymath}
\citep[equation~(3.35)]{B}, where $r$ is the distance traveled along the streamline parallel to $\mathbf{q}_{tot}$ and defined as $dr = \left| \mathbf{q}_{tot}\right| d\tau$ (see \autoref{fig:streamline}). 
Note that this leads to $d\tau/dr = 1/\left| \mathbf{q}_{tot}\right|$, which is the transformation proposed by \citep[equation~(15)]{ZT}. We can further define the volumetric flow rate $q_{sl, tot}$ along a streamline by 
\begin{equation}
\label{eq:Aq}
	q_{sl, tot} = \left| \mathbf{q}_{tot}\right| A(r),
\end{equation}
where $A(r)$ is the cross-section of a streamtube bounded by streamlines (see \autoref{fig:streamline}). By definition of a streamtube, we have the fundamental property,  
\begin{displaymath}
	\frac{\partial q_{sl, tot}}{\partial r} = 0,
\end{displaymath}
which is a restatement of \autoref{eq:incompressible}. This means that it is possible, using a parameterization along streamlines, 
to reduce \autoref{eq:consmass} to a sum of independent 1D problems along streamlines by the change of variables $S_w(t, \vec{x}) = \tilde{S}_w(t, r)$, 
\begin{equation}
	\label{eq:cons1}
	\phi(r) A(r) \frac{\partial \tilde{S}_w}{\partial t} + q_{sl, tot}(t) \frac{\partial f_w(\tilde{S}_w)}{\partial r} = 0.
\end{equation}

\begin{figure}
	\begin{subfigure}[b]{.45\textwidth}
		\begin{tikzpicture}[scale=2.5,>=to,line width=1.5pt,every text node part/.style={align=center}]
\draw [->](0,0) -- (0,1); 
\draw (2,-0.1) node(text1){$Z$};

\draw [->](0,0) -- (2,0); 
\draw (-0.1,1) node(text2){$S_w$};

\draw[dotted] (0,0.1) -- (2,0.1); 
\draw (-0.1,0.1) node(text3){$s_{wi}$};

\draw[dotted] (0,0.45) -- (2,0.45); 
\draw (-0.1,0.45) node(text4){$s^*$};

\draw[dotted] (0,0.9) -- (2,0.9); 
\draw (-0.1,0.9) node(text5){$s_B$};

\draw[dotted] (0.12, 0) -- (0.12, 1); 
\draw (0.12,-0.1) node(text6){$f_w'(s_B)$};

\draw[dotted] (1.4, 0) -- (1.4, 1);  
\draw (1.4,-0.1) node(text7){$\alpha^*$};

\draw[red, thick] (0, 0.9) -- (0.12, 0.9); 
\draw[red, thick] (0.12, 0.9) to [out=-40, in=170] (1.4, 0.45); 
\draw[red, thick] (1.4, 0.45) -- (1.4, 0.1); 
\draw[red, thick] (1.4, 0.1) -- (2, 0.1); 


\draw (1.3,1) node(text8){$s_w$};
\draw [->,shorten >=5pt,gray] (text8)  to [out=190,in=30] (0.8,0.53);

\end{tikzpicture}
	\caption{Sketch of the solution mapping $\check{S}_w(Z)$}
	\label{fig:solBL}
	\end{subfigure}
	\begin{subfigure}[b]{.45\textwidth}
		\begin{tikzpicture}[scale=2.5,>=to,line width=1.5pt,every text node part/.style={align=center}]

\draw [->](0,0) -- (1,0); 
\draw (1.2,-0.1) node(text2){$S_w$};

\draw [->](0,0) -- (0,2); 
\draw (-0.1,2) node(text1){$Z$};

\draw[dotted] (0.1,0) -- (0.1,2); 
\draw (0.1,-0.1) node(text3){$s_{wi}$};

\draw[dotted] (0.45,0) -- (0.45,2); 
\draw (0.45,-0.1) node(text4){$s^*$};

\draw[dotted] (0.9,0) -- (0.9,2); 
\draw (0.9,-0.1) node(text5){$s_B$};

\draw[dotted] (0,0.12) -- (1,0.12); 
\draw (-0.2,0.12) node(text6){$f_w'(s_B)$};

\draw[dotted] (0,1.4) -- (1,1.4);  
\draw (-0.1,1.4) node(text7){$\alpha^*$};

\draw[red, thick] (0.9,0.12) to [out=130, in=-80] (0.45,1.4); 

\draw[red, thick] (0.45,1.4) -- (0.1,1.4); 


\draw (1.3,1.1) node(text8){$s_w^{<-1>} = f_w'$};
\draw [->,shorten >=5pt,gray] (text8)  to [out=230,in=0] (0.6,0.8);

\end{tikzpicture}
	\caption{Sketch of the inverse mapping of $\check{S}_w(Z)$}
	\label{fig:solBLinv}
	\end{subfigure}
	\caption{Snapshot of the solution $\check{S}_w(Z)$ (\ref{fig:solBL}) and its inverse $\check{S}^{<-1>}_w(s)$ (reproduction from [fig.1 in \cite{IMT}]) (\ref{fig:solBLinv})}
\end{figure}
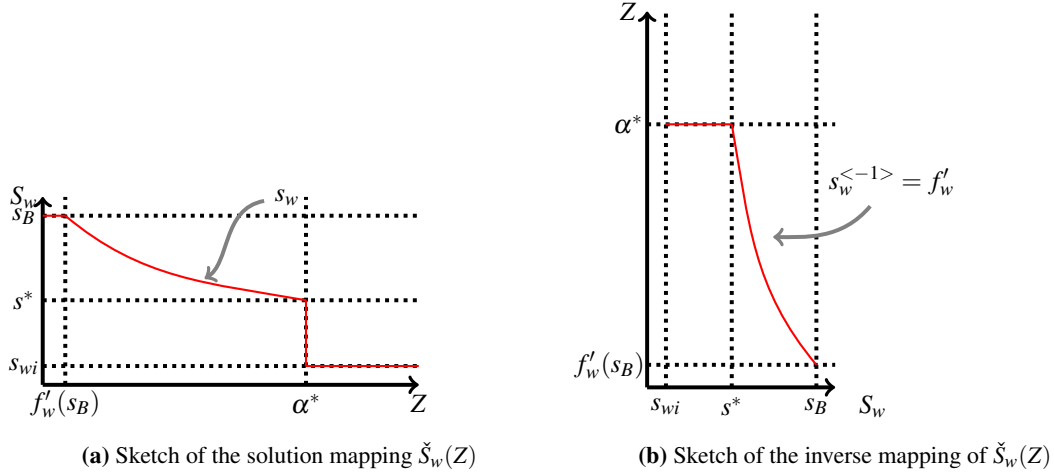

Finally, following \cite{HY} by introducing the cumulative injection volume $Q$ and the cumulative pore volume $V$ per streamline, 
\begin{equation}
\label{eq:cumu}
		Q(t) = \int_0^t q_{sl, tot}(t') dt' \mbox{ and }
		V(r) = \int_0^r \phi(r') A(r') dr',
\end{equation}
the change of variables $\tilde{S}_w(t, r) = \check{S}_w(Q, V)$ reduces \autoref{eq:cons1} to
\begin{equation}
	\label{eq:cons2}
	\frac{\partial \check{S}_w}{\partial Q} + \frac{\partial f_w(\check{S}_w)}{\partial V} = 0
\end{equation}
together with the initial and boundary conditions, 
\begin{equation}
\label{eq:cons2IBC}
\check{S}_w(0, V) = s_{wi} \mbox{ and }
\check{S}_w(Q, 0) =1 - s_{oi} = s_B,
\end{equation}
respectively, where $s_{wi}$ is the initial water saturation of the reservoir and $s_{B}$ is the water injection (both assumed to be known, and uniform and constant, respectively).
Hence, $\check{S}_w: \; (Q,V) \mapsto \check{S}_w(Q, V)$ is a deterministic mapping from the uncertain reservoir characteristics (permeability and porosity) to the 
random water saturation field. The key advantage of this reformulation lies in the fact that the mapping equation (along each streamline) is a 1D equation in the stochastic variables, and can therefore be solved efficiently \citep{IMT}.  
Moreover, the mapping \autoref{eq:cons2} admits an analytical solution in terms of the ratio $Z$ between $V$ and $Q$ \citep{HY}, simplifying even further the interpretation of the mapping function,
\begin{equation}
	\label{eq:solBL}
	\check{S}_w\left(Z = \frac{V}{Q}\right) = 
	\left\{
	\begin{array}{lll}
		s_B & \mbox{if} & Z < f_w'(s_B) \\
		s_w\left(Z\right) & \mbox{if} & f_w'(s_B) \leq Z < f_w'(s^*) \\
		s_{wi} & \mbox{if} & Z \geq f_w'(s^*), \\
	\end{array}	
	\right.
\end{equation}
where $s^*$ is defined by the Rankine-Hugoniot condition
\begin{displaymath}
f'_w(s^*) = \frac{f_w(s^*) - f_w(s_{wi})}{s^* - s_{wi}},
\end{displaymath}
$s_B$ is defined in \autoref{eq:cons2IBC}, and $s_w(y)$ is defined for $y < f'_w(s^*)$ as $s_w(y) = (f'_w)^{-1}(y)$. Let us call $\alpha^* = f'_w(s^*)$. This self-similarity behavior sketched in \autoref{fig:solBL} is a key factor in our method, as this leads to a tremendous reduction in the complexity of the stochastic problem. Indeed, even if the permeability field is a full stochastic tensor, $Z$ is always a stochastic scalar field. Once the deterministic water saturation mapping $\check{S}_w$ is solved, the uncertainties in the input parameters $V$ and $Q$ translate through the mapping or its inverse to the uncertainty in $S_w$. The mapping equation shows that physically, propagating the uncertainties in the porosity and permeability fields is equivalent to propagating the uncertainties in $Z$ along (uncertain) streamlines.

\medskip

Because the saturation is defined by a nonlinear equation (\autoref{eq:consmass}), the stochastic saturation field $S_w$ may be elaborate (e.g. far from Gaussian). Therefore, instead of working directly with $S_w$ as a stochastic field, we regard the saturation as a nonlinear mapping function $\check{S}_w(Z)$ of "smoother" stochastic fields that constitute $Z$. This mapping function, deterministic and either analytically (in \autoref{eq:solBL}) or numerically (from \autoref{eq:cons2}-\autoref{eq:cons2IBC}) determined, entirely embodies the nonlinearity of the saturation. The "smooth" stochastic fields drive the probabilistic distribution of the saturation. This reformulation was used in \citep{IMT} for estimating the saturation distribution in the spatial 1D case and is extended to the multi-dimensional case in this section. Indeed, $V$ and $Q$ are defined for each streamline and cannot be used in a computationally efficient manner to estimate the distribution of $Z$ in multiple spatial dimensions since the streamline paths are time dependent and also stochastic. We are going to rewrite $Z$ into a ratio of quantities that can be more easily and efficiently estimated. 
Along a streamline $\vec{x}_{sl}(\tau)$, as pictured in \autoref{fig:streamline}, the water saturation is 
\begin{equation}
\label{eq:paramS}
	S_w\left[\vec{x} = \vec{x}_{sl}(\tau(\vec{x})), t\right] = \check{S}_w\left[Z(r(\tau(\vec{x})),t)\right], 
\end{equation}
with 
$\vec{x}_{sl}(\tau)$ defined in \autoref{eq:cons} and
\begin{equation}
\label{eq:curvesl}
r(\tau) = \int_0^{\tau} \left| \mathbf{q}_{tot}\left[\vec{x}_{sl}(t'), t'\right] \right| dt' = \int_0^{\tau} q_{tot}\left[\vec{x}_{sl}(t'), t'\right] dt'
\end{equation}
being the streamline travel distance to position $\vec{x}$ and $Z(r,t) = V(r)/Q(t)$, with $Q(t)$ and $V(r)$ defined in \autoref{eq:cumu}. More precisely, we should write $Q^{sl(t)}(t)$ and $V^{sl(t)}(r)$ as the streamlines are also evolving in time. This last remark highlights why it is difficult to efficiently estimate the distributions of these two quantites in multidimensional porous media as these distributions depend on the streamline geometry and saturation distribution. To alleviate this difficulty, we make a key approximation: the frozen streamline approximation.

\subsection{Saturation distribution estimation via stochastic travel and injection time}
\label{subsection:FROST}

We assume that the streamline pattern is dominated by the permeability heterogeneity and is only weakly influenced by local viscosity effects due to the wetting phase displacing the non-wetting phase. 
Within this frozen streamline approximation, $\mathbf{q}_{tot}\left(\vec{x}, t\right)$ can be written as 
\begin{equation}
\label{eq:fixedsl}
	\mathbf{q}_{tot}\left(\vec{x}, t\right) = q_{tot}\left(\vec{x}, t\right) \mathbf{e}_{q}(\vec{x}),
\end{equation}
where $\mathbf{e}_{q}(\vec{x})$ is a time-independent unity direction vector parallel to $\mathbf{q}_{tot}$. We introduce $\vec{x}_{sl}(\tau) = \vec{\bar{x}}_{sl}\left[r(\tau)\right]$, so that $\vec{\bar{x}}_{sl}$ is a function of $r$ only. We can then write 
\begin{equation}
\label{eq:deriveSLposition}
    \frac{d \vec{x}_{sl}}{d\tau} = \frac{d\vec{\bar{x}}_{sl}}{dr} \frac{dr}{d\tau} = \frac{d\vec{\bar{x}}_{sl}}{dr} q_{tot},
\end{equation}
where the second equality results from \autoref{eq:curvesl}, and from \autoref{eq:cons} and \autoref{eq:fixedsl}
\begin{equation}
\label{eq:deriveSLposition2}
	\frac{d \vec{x}_{sl}}{d\tau} = \mathbf{q}_{tot} = q_{tot} \mathbf{e}_q.
\end{equation}
From \autoref{eq:deriveSLposition} and \autoref{eq:deriveSLposition2}, we can derive the following streamline-generating relation,   
\begin{equation}
	\frac{d \vec{\bar{x}}_{sl}}{dr} = \mathbf{e}_q\left(\vec{\bar{x}}_{sl}\right). 
\end{equation}
Hence, under the fixed or frozen streamline approximation, the streamline pattern is time-invariant, $sl(t) = sl(0)$, and the streamtube cross-sectional areas do not change in time $A(r;t) = A(r)$. Therefore, $V$ only depends on the streamline distance from the injection, $V(r;t) = V(r)$, and $Z(r,t)$ is a separable function in $r$ and $t$.

The frozen streamline approximation, though simplistic, is well suited for highly heterogeneous permeability fields \citep{ZT,MJM}, which are typically the case in applications. Therefore the following framework is remarkably appropriate for high input variances $\sigma_K^2$.

Based on \autoref{eq:Aq}, the cross-section area at $r(\tau(\vec{x}))$ along the streamtube can be expressed as 
\begin{equation}
\label{eq:crossA}
	A[r(\tau(\vec{x}))] = \frac{q_{sl,tot}(t; \vec{x})}{q_{tot}[r(\tau(\vec{x})),t]}. 
\end{equation}
Within the frozen streamline approximation, $A[r(\tau(\vec{x}))]$ is time-independent and thus can be determined for example by $\mathbf{q}_{tot}$ at time $t=0$. Therefore, for each spatial location $\vec{x}$ in the reservoir, we have from \autoref{eq:cumu}
\begin{equation}
\label{eq:Z}
	Z(\vec{x},t) = \frac{V(\vec{x})}{Q(\vec{x},t)} = \dfrac{\ds\int_0^{r(\tau(\vec{x}))} \dfrac{\phi(r')}{q_{tot}(r', t=0)} dr'}{\ds\int_0^t \dfrac{q_{sl,tot}(t'; \vec{x})}{q_{sl,tot}(t=0; \vec{x})} dt'}. 
\end{equation}
The denominator $\int_0^t q_{sl,tot}(t')/q_{sl,tot}(t=0) dt'$ will thereafter be referred to as the Equivalent Injection Time (EIT) as it corresponds to the time needed to inject at the rate $q_{sl,tot}(t=0)$ a volume equivalent to $\int_0^t q_{sl,tot}(t') dt'$. The volumetric flow rate $q_{sl,tot}$ is an integral quantity of a streamtube, i.e. $q_{sl,tot} = \Delta p/R$, with resistance $R = \int_0^L [A(r')\lambda_{tot}(r')]^{-1} dr'$ and pressure drop $\Delta p$. Correspondingly, we may further approximate the EIT to be an almost deterministic time-dependent function that can be estimated in terms of a mean value, $\langle EIT \rangle$, from few MCS samples. The scope of this EIT approximation is studied in \autoref{section : companal}. In the case of rate control boundary conditions, $q_{sl,tot}(t;\mathbf{x})$ and consequently EIT may become a deterministic quantity.

The numerator $\int_0^r \phi(r')/q_{tot}(r',t=0) dr'$, on the other hand, is exactly the TOF $\tau$ for the initial streamline pattern. The TOF captures both uncertainties in the porosity and permeability fields. Its distribution can be estimated with small computational cost, since no transport sampling is needed. 
To estimate the distribution of the TOF, we generate realizations of streamlines that go through every point $\vec{x}$ of interest using Pollock's algorithm \citep{Pol,HY}. This is accomplished by backtracking streamlines in time from $\vec{x}$ to the injection location. The resulting statistics of the TOF are needed to determine the distribution of $Z(\vec{x},t)$ by means of kernel density estimation (KDE) \citep{BGK}. 
The TOF can also be estimated by solving an elliptic steady-state equation in Eulerian coordinates \citep{SMWG}.

Similarly, thanks to the frozen streamline approximation and \autoref{eq:crossA}, the EIT can be approximated as follows, 
\begin{equation}
\label{eq:EIT_simplified}
	\begin{array}{rl}
		EIT(\vec{x},t) & = \ds \int_0^t \frac{q_{tot}[r(\tau(\vec{x})),t'] A[r(\tau(\vec{x}))]}{q_{tot}[r(\tau(\vec{x})),t=0] A[r(\tau(\vec{x}))]} dt', \\
		&= \ds \int_0^t \frac{q_{tot}[r(\tau(\vec{x})),t']}{q_{tot}[r(\tau(\vec{x})),t=0]} dt', \\
		&= \ds \int_0^t \frac{q_{tot}(\vec{x},t')}{q_{tot}(\vec{x},t=0)} dt', \\
	\end{array}
\end{equation}
which requires the total velocity field, but not the construction of streamlines.

\subsection{Case-specific simplifying approximations}
\label{subsection:Simplific}

The TOF is strongly linked to the velocity field, which is described by a linear equation when the saturation is fixed. If the permeability field has a `smooth' distribution, we expect the TOF to have a similar behavior. In the cases studies in this work, the permeability field is assumed to be lognormal, so that the logarithm of the TOF is the appropriate stochastic field to be considered. 
We therefore approximate $Z$ to be 
\begin{equation}
\label{eq:approxZ}
	Z = \dfrac{\exp\left(\log\tau\right)}{\langle EIT\rangle}
\end{equation}
and estimate the CDF of the water-saturation at position $\vec{x}$ and time $t$, as follows, 
\begin{eqnarray}
\label{eq:multiDCDFsaturation}
\lefteqn{F^{FROST}_{S_w}(s; \vec{x},t) = \mathcal{P}\left\{ S_w\left(\vec{x},t\right) \leq s \right\}} \nonumber \\
			& = & \mathcal{P}\left\{ \check{S}_w\left[Z(\vec{x},t)\right] \leq s \right\} \nonumber \\
			& = & \mathcal{P}\left\{ Z(\vec{x},t) \geq \chi_F(s) \right\} \nonumber \\
			& = & 1 - \mathcal{P}\left\{ Z(\vec{x},t) \leq \chi_F(s) \right\} \\
			& = & 1 - \mathcal{P}\left\{ e^{\log\tau(\vec{x})} / \langle EIT\rangle(\vec{x},t) \leq \chi_F(s) \right\} \nonumber \\
			& = & 1 - \mathcal{P}\left\{ \log\tau(\vec{x}) \leq \log\left[\chi_F(s) \langle EIT\rangle(\vec{x},t) \right] \right\} \nonumber \\
			& = & 1-F_{\log\tau(\vec{x})}\left[\log\left(\chi_F(s) \langle EIT \rangle (\vec{x},t) \right)\right], \nonumber
\end{eqnarray}
where $F^{FROST}_{S_w}(s; \vec{x},t)$ is the FROST estimate of the saturation CDF at saturation level $s$, $F_{\log \tau(\vec{x})}$ is the $\log$(TOF) CDF at position $\vec{x}$, $\mathcal{P}$ is the probability measure, and $\chi_F(s)$ corresponds to the inverse mapping $\check{S}^{<-1>}_w$ illustrated in \autoref{fig:solBLinv}. The latter is defined as follows, 
\begin{displaymath}
\begin{array}{rl}
	\chi_F(s) & = 
		\left\{
		\begin{array}{ll}
			+\infty, & s < s_{wi}, \\
			\alpha^* , & s \in (s_{wi}, s^*), \\
			f_w'(s), & s \in (s^*, s_B), \\
			0, & s > s_B.
		\end{array}
		\right.
\end{array}
\end{displaymath}
The derivation of \autoref{eq:multiDCDFsaturation} is deduced from \autoref{eq:approxZ},
 \autoref{eq:paramS} that gives the mapping from $Z$ to $S_w$, and the analytical expression of the mapping in \autoref{eq:solBL} . The saturation PDF is given by taking the derivative of \autoref{eq:multiDCDFsaturation} with respect to $s$, 
\begin{equation}
\begin{array}{rl}
\label{eq:multiDPDFsaturation}
	p_{S_w}^{FROST}(s; \vec{x},t) & = F^*_{S_w}(\vec{x},t) \delta_0(s) - \pi(s;\vec{x},t), \mbox{ with} \\
	F^*_{S_w}(\vec{x},t) & = 1-F_{\log\tau(\vec{x})}\left[\log\left(\alpha^* \langle EIT \rangle (\vec{x},t) \right)\right], \\
	\pi(s;\vec{x},t) & = \dfrac{f_w''(s)}{f_w'(s)}p_{\log\tau(\vec{x})}\left[\log\left(\chi_p(s) \langle EIT \rangle (\vec{x},t) \right)\right],
\end{array}
\end{equation}
$\delta_0$ being the Dirac measure, and
\begin{displaymath}
\label{eq:Chip}
\begin{array}{rl}
	\chi_p(s) &= 
		\left\{
		\begin{array}{ll}
			f_w'(s), & s \in (s^*, s_B), \\
			0, & \mbox{otherwise}. 
		\end{array}
		\right.
\end{array}
\end{displaymath}

In \autoref{subsection:EIT} we argue that, for time-independent pressure control, the mean EIT approximately follows a power law and is only a function of time (and implicitly, of the viscosity ratio $m$), 
\begin{equation}
\label{eq:meanEIT_power}
	\langle EIT \rangle (\vec{x},t) \approx c(m) t^{\beta(m)}, 
\end{equation}
where $c(m)$ and $\beta(m)$ are estimated using two time steps at times $t = \Delta t$ and $2 \Delta t$ as 
\begin{equation}
\begin{array}{rl}
\label{eq:meanEIT_parameters}
c & = \langle EIT \rangle (\vec{x},\Delta t) / \Delta t^{\beta} \mbox{ and} \\
\beta & = \ds \left.\log\left[\frac{\langle EIT \rangle (\vec{x},2\Delta t)}{\langle EIT \rangle (\vec{x},\Delta t)}\right] \right/ \log(2)
\end{array}
\end{equation}
for any $\vec{x}$. For viscosity ratios $m < 1$ with less viscous water relocating more viscous oil, the volumetric flux is expected to increase with time, so that $\beta(m) > 1$. For $m$ close to unity, the mean EIT scales linearly with $t$ as $\beta(m) \approx 1$, and for $m  > 1$, we expect $\beta(m) < 1$.  

Therefore, under the EIT approximation made in \autoref{eq:meanEIT_power}, the simplified FROST saturation CDF in \autoref{eq:multiDCDFsaturation} ultimately becomes 
\begin{equation}
\label{eq:multiDCDFm1}
	F^{FROST}_{S_w}(s; \vec{x},t) = 1-F_{\log\tau(\vec{x})}\left[\log\left(\chi_F(s) c(m)t^{\beta(m)} \right)\right].
\end{equation}

Besides the frozen streamline approximation, \autoref{eq:multiDCDFsaturation} and \autoref{eq:multiDCDFm1} are valid if the EIT has a small variance so that we can replace it by its average. This constraint can restrict the range of possible viscosity ratios $m$. We investigate the effect of $m$ on the EIT variance in \autoref{subsection:EIT}.
To summarize, the FROST saturation one-point distribution estimate solely depends on two key statistics: the logarithm of the TOF field ($\log$(TOF)), $\log\tau(\vec{x})$ and the average EIT, $\langle EIT \rangle (\vec{x},t)$. Once these quantities are obtained, the saturation distribution is readily available through \autoref{eq:multiDCDFsaturation}. Therefore, the efficiency and accuracy of the FROST distribution method strongly depends on the computational costs of generating accurate estimates of these two statistics. The mean EIT has a parametric formula (see \autoref{eq:meanEIT_power} and \autoref{eq:meanEIT_parameters}) that a priori involves solving the flow and transport problem for two time steps, and the $\log$(TOF) distribution can be evaluated using KDE on samples at time $t=0$. Alternatively, for example under conditions studied in \autoref{section : companal}, a parametric PDF model or surrogate for the $\log$(TOF) may be applied.

	\subsection{The FROST algorithm}
	\label{subsection:FROSTalgo}

	\autoref{eq:multiDCDFm1} provides a closed-form, semi-analytical method to estimate the saturation distribution that relies on the distribution of the $\log$(TOF), computed from a linear problem, and on the mean EIT that can be estimated by solving the two-phase flow problem for two time steps. We verify that this formula produces accurate estimates of the saturation distribution. The FROST algorithm is exposed in Algorithm \ref{alg:DM}, where $N_s$ is the number of initial travel time realizations, $M_s$ is the number of times we solve the transport problem for two times steps, and $N_t$ is the number of time steps considered during the entire simulation.

		\begin{algorithm}
		\begin{algorithmic}
			\For {$i = 1\cdots N_s$}
				\State Solve initial tracer equation 
				\State Store time-of-flights $\tau(\vec{x})^{(i)}$
                \If {$i \leq M_s$}
                	\State Solve transport eq. at times $T/N_t$ and $2T/N_t$
                	\State Store velocity fields $\left\{q^{(i)}_{tot}(kT/N_t)\right\}_{k=1,2}$
                 \EndIf
			\EndFor
			\State Estimate $p_{\log\tau(\vec{x})}$ and $F_{\log\tau(\vec{x})}$ with KDE using $\{ \tau(\vec{x})^{(i)}\}_{1\leq i \leq N_s}$
			\State Estimate $c$, $\beta$ and $\langle EIT \rangle (t) = c t^\beta$ using $\left\{q^{(i)}_{tot}(kT/N_t)\right\}_{k=1,2; 1\leq i \leq N_s}$
			\State Compute $\left\{F^{FROST}_{S_w}(s; \vec{x},t_j)\right\}_{j = 1 \dots N_t}$ and $\left\{p^{FROST}_{S_w}(s; \vec{x}, t_j)\right\}_{j = 1 \dots N_t}$ for $t_j = j T/N_t$ 		
			\end{algorithmic} 
		\caption{Frozen streamline-based distribution method (FROST)}
		\label{alg:DM}
		\end{algorithm}

	\subsection{Special case: tracer concentration}
	\label{subsection : tracer}
	
		Before focusing on two-phase flow, we explore the implications of the previously derived distribution method in the case of advection-dominated tracer dispersion. The hydraulic conductivity field is uncertain and assumed to be modeled as a stochastic field with known (log Gaussian) distribution. If $C(\vec{x},t)$ denotes the tracer concentration at position $\vec{x}$ and time $t$, we are interested in the following stochastic problem, 
		\begin{displaymath}
		\left\{ 
			\begin{array}{ll}
				\ds \frac{\partial C}{\partial t} + \nabla \cdot \left(\mathbf{u} C\right) = 0, \\
				\nabla \cdot \left(K \nabla h\right) = 0, \\
				\mathbf{q} = -K \nabla h, \; \mathbf{u} = \mathbf{q}/n, \\
				h(\vec{x},t) = h_{\mbox{\small{in}}}, C(\vec{x},t) = C_0, \; \vec{x} \in \Gamma_{inj}, \; \forall t \\
				h(\vec{x},t) = h_{\mbox{\small{out}}}, \; \vec{x} \in \Gamma_{prod}, \; \forall t \\
				C(\vec{x},0) = 0, \; \vec{x} \notin \Gamma_{inj},
			\end{array}
		\right.
		\end{displaymath} 
where $K$ is the hydraulic conductivity field, $h$ is the hydraulic head, $n$ is the porosity field, $\mathbf{q}$ is the specific discharge and $\mathbf{u}$ is the velocity field. Since the conductivity field is random, so is the hydraulic head, and the specific discharge by construction. The velocity field inherits randomness possibly from both the specific discharge and the porosity field. And finally the tracer concentration becomes random from transport with random velocity. 
	
	The tracer case can be deduced from the two-phase flow case. Indeed, this case corresponds to the situation where the `fractional' flow function is the identity $f_w(s) = s$. In this case, no shocks are developed during the transport. Furthermore, since we only have one phase, the EIT is deterministic and $EIT(\vec{x},t) = t$. Using the distribution method, we deduce the following formula for the expected tracer concentration, 
	\begin{displaymath}
	\begin{array}{rl}
		\langle C \rangle (\vec{x},t) &= C_0 \Pb\left( Z < 1\right) \\
		&= C_0 F_{\log\tau(\vec{x})}\left[\log\left(t \right)\right]
	\end{array}
	\end{displaymath}
	The tracer concentration variance is also deduced from the CDF of the logarithm of the TOF, 
	\begin{displaymath}
	\begin{array}{rl}
		\sigma_C^2 (\vec{x},t) &= C_0^2 F_{\log\tau(\vec{x})}\left[\log\left(t \right)\right] 
		\left(1 - F_{\log\tau(\vec{x})}\left[\log\left(t \right)\right] \right).
	\end{array}
	\end{displaymath}

\section{Numerical studies of the FROST approximations and performance}
\label{section : companal}

\begin{figure}
	                \begin{tikzpicture}[scale=3.5,>=to,line width=1.5pt,every text node part/.style={align=center}]

\draw [->](0,0) -- (0,1); 
\draw [->](0,0) -- (1,0); 
\draw [-](0,1) -- (1,1); 
\draw [-](1,0) -- (1,1); 

\draw [] (1,-.1) node(bla1){$x_1$};
\draw [] (-.1,1) node(bla2){$x_2$};

\draw [-,very thick,blue](0,0) -- (1/7,0);
\draw [->,blue] (1/14,-1/7) -- (1/14,-1/20);
\draw [blue] (2/7,-1.5/14) node(bla){injector};

\draw [-,very thick,red](1-1/7,1) -- (1,1);
\draw [->,red, very thick] (1-1/14,1/20+1) -- (1-1/14,1/7+1);
\draw [red] (1-2/7,1+1.25/14) node(bla2){producer};

\draw[gray] [-](1/3,0) -- (1/3,1); 
\draw[gray] [-](2/3,0) -- (2/3,1); 
\draw[gray] [-](0,1/3) -- (1,1/3); 
\draw[gray] [-](0,2/3) -- (1,2/3); 

\draw[gray, dotted, thin] [-](1/6,0) -- (1/6,1); 
\draw[gray, dotted, thin] [-](3/6,0) -- (3/6,1); 
\draw[gray, dotted, thin] [-](5/6,0) -- (5/6,1); 
\draw[gray, dotted, thin] [-](0,1/6) -- (1,1/6); 
\draw[gray, dotted, thin] [-](0,3/6) -- (1,3/6); 
\draw[gray, dotted, thin] [-](0,5/6) -- (1,5/6); 

\draw[red] (1/6,1/6) node{\textbullet};
\draw[red] (1/6,3/6) node{\textbullet};
\draw[red] (1/6,5/6) node{\textbullet};
\draw[red] (3/6,1/6) node{\textbullet};
\draw[red] (3/6,3/6) node{\textbullet};
\draw[red] (3/6,5/6) node{\textbullet};
\draw[red] (5/6,1/6) node{\textbullet};
\draw[red] (5/6,3/6) node{\textbullet};
\draw[red] (5/6,5/6) node{\textbullet};

\draw [red, thick] (1/6,1/6-1/12) node(bla){$(1,1)$};
\draw [red, thick] (3/6,1/6-1/12) node(bla){$(2,1)$};
\draw [red, thick] (5/6,1/6-1/12) node(bla){$(3,1)$};
\draw [red, thick] (1/6,3/6-1/12) node(bla){$(1,2)$};
\draw [red, thick] (3/6,3/6-1/12) node(bla){$(2,2)$};
\draw [red, thick] (5/6,3/6-1/12) node(bla){$(3,2)$};
\draw [red, thick] (1/6,5/6-1/12) node(bla){$(1,3)$};
\draw [red, thick] (3/6,5/6-1/12) node(bla){$(2,3)$};
\draw [red, thick] (5/6,5/6-1/12) node(bla){$(3,3)$};

\end{tikzpicture}
	                \caption{Computational quarter-five spot setting. The solid blue line South-West corresponds to the injection spot (e.g. contaminant or water). The solid red line North-East corresponds to the pumping (or production) well. The red dots in the porous medium corresponds to locations where all parameters (e.g. TOF, EIT, PDF and CDF) are investigated.}
	                \label{fig:45spot}
	\end{figure}
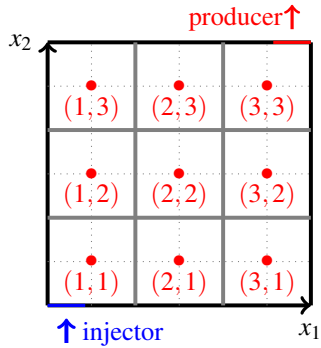

Throughout the rest of the manuscript, we perform tests on a 2D quarter-five spot configuration. This configuration corresponds to a squared porous medium where fluid is injected in the south-west corner and a pumping well is located in the north-east corner (see \autoref{fig:45spot}). We set the dimensions of the porous domain as $L_x = L_y = L = 1$. The typical grid used for simulations is $128 \times 128$. Although we measure the statistics of interest (TOF, EIT and saturation) on the entire grid, we illustrate the results on 9 spots, labeled $(i,j), \; 1 \leq i,j \leq 3$ as depicted in \autoref{fig:45spot}. 
In the rest of the manuscript, we analyze the two-phase flow in the squared porous medium modeled as in \autoref{eq:incompressible} and \autoref{eq:consmass},
with pressure control dictated by \autoref{eq:BCpressure}, under stochastic rock properties. We solve the two-phase flow problem for a number of realizations, $N_s$, with pressure control defined in \autoref{eq:BCpressure} and $\Gamma_{\mbox{\small{inj}}}$ and $\Gamma_{\mbox{\small{prod}}}$ spanning on $4$ grid segments, $p_{inj} = 8$ and $p_{prod} = 0$ (see \autoref{fig:45spot}). Unless otherwise stated, the final simulation time is $T=2$. 

In this section, we assume that $K$ is a stochastic field with $\log(K)$ being a Gaussian field with exponential covariance, while $\phi = 0.3$ is assumed to be a deterministic constant. More precisely, $\log K \sim \mathcal{N}(\mathbf{0}, C_K)$, where 
		\begin{displaymath}
			C_K(\vec{x}, \vec{y}) = \sigma_K^2 \exp\left(- \sqrt{\sum_{i=1}^d \dfrac{(x_i - y_i)^2}{\lambda_i^2}}\right), 
		\end{displaymath}
		with $\sigma_K^2$ being the variance of $\log(K)$, $\lambda_i$ the correlation length in each dimension and $d = 2$ the physical spatial dimension. The effects of stochasticity in $\phi$ were partially studied in \citep{IMT} and are deferred to future work as they only affect the distribution of log(TOF).
		We use the streamline solver from \citep{MJM} to solve for all quantities of interest.

The goal of this section is to evaluate the performance of the FROST method for log-Gaussian permeability fields and assess the accuracy of the aforementioned approximations and metrics used to derive saturation statistics. In subsection \ref{subsection:smoothness}, we verify that, in this setting, the TOF distribution is indeed smooth and can be approximated by a log-Gaussian field. In subsection \ref{subsection:accuracy}, we compare saturation CDF and PDF estimates from both MCS and FROST. We argue that the FROST method produces accurate results compared to MCS despite the frozen streamline approximation. Furthermore, the FROST saturation PDF estimation is more stable than the KDE-based MCS results, especially in areas where the injected water front is likely to be located. Finally we verify that, fixing the initial streamlines in MCS, the frozen streamline approximation generates exactly matching saturation CDFs between the FROST and MCS methods. In light of the observations from the latter subsections, we propose, in subsection \ref{subsection:efficiency}, a computationally accelerated version of FROST based on a surrogate Gaussian distribution for the log(TOF). Finally, in subsection~\ref{subsection:EIT}, we evaluate the scope of the EIT approximation proposed in \autoref{eq:meanEIT_power}. 
		
	\subsection{Smoothness of $\log$(TOF) distribution}
	\label{subsection:smoothness}
	
\begin{figure*}
					\centering
					\includegraphics[scale=0.7]{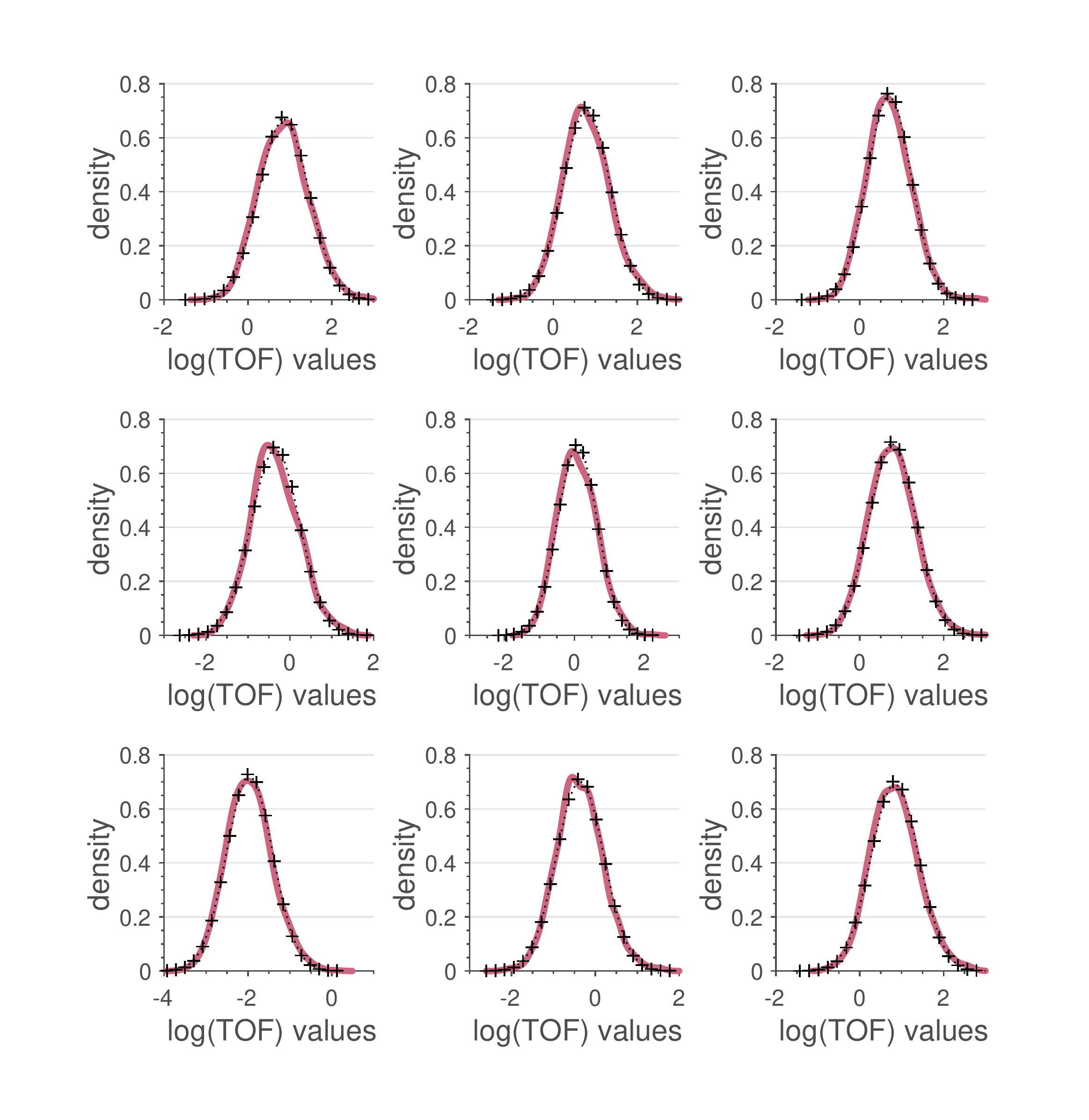}		
	                \caption{Illustration of smoothness of $\log$(TOF) in the case of log Gaussian permeability field at the nine locations depicted in \autoref{fig:45spot}. The permeability field distribution has a uniform correlation length $\lambda = 0.1L$, variance $\sigma_K^2 = 1$ and the viscosity ratio is set to be $m = 0.25$. $N_s = 2000$ realizations were used to generate the PDFs. The solid lines are estimations of the $\log$(TOF) PDF with KDE, while the pluses correspond to a Gaussian fitting using sample means and standard deviations.}
	                \label{fig:logTOF_pdfs1}
	\end{figure*}

We record the TOF using Pollock's algorithm (\cite{Pol}). We then use state-of-the-art KDE \citep{BGK} to estimate the distribution of the $\log$(TOF). With $\sigma_K^2 = 1$, $\lambda_1 = \lambda_2 = \lambda = 0.1L$, $m = 0.25$ and $N_s = 2000$, PDFs of the log(TOF), $p_{\log\tau(\vec{x})}$, at the nine locations shown in \autoref{fig:45spot}, are displayed in \autoref{fig:logTOF_pdfs1}. We can see that at all nine observation spots (and really in the entire grid), $p_{\log \tau(\vec{x})}$ is smooth and close to a Gaussian surrogate, $p_{\mathcal{G}(\log \tau(\vec{x}))}$, simply built as follows, 
\begin{displaymath}
	p_{\mathcal{G}(\log \tau(\vec{x}))} \sim \mathcal{N}(m^{(N_s)}(\vec{x}), (s^{(N_s)}(\vec{x}))^2)
\end{displaymath}
with sample mean
\begin{displaymath}
m^{(N_s)}(\vec{x}) = \frac{1}{N_s} \sum_{k = 1}^{N_s} \log \tau(\vec{x})^{(k)}
\end{displaymath}
and sample variance
\begin{displaymath}
(s^{(N_s)}(\vec{x}))^2 =  \frac{1}{N_s - 1} \sum_{k = 1}^{N_s} \left( \log \tau(\vec{x})^{(k)} - m^{(N_s)}(\vec{x}) \right)^2
\end{displaymath}
based on available realizations $[ \log \tau(\vec{x})^{(k)} ]_{1 \leq k \leq N_s}$ of the $\log$(TOF). 
		This illustrates that under certain conditions the generation of the $\log$(TOF) distribution can be further accelerated by means of Gaussian surrogates.

\begin{figure}
					\centering
					\includegraphics[scale=0.4]{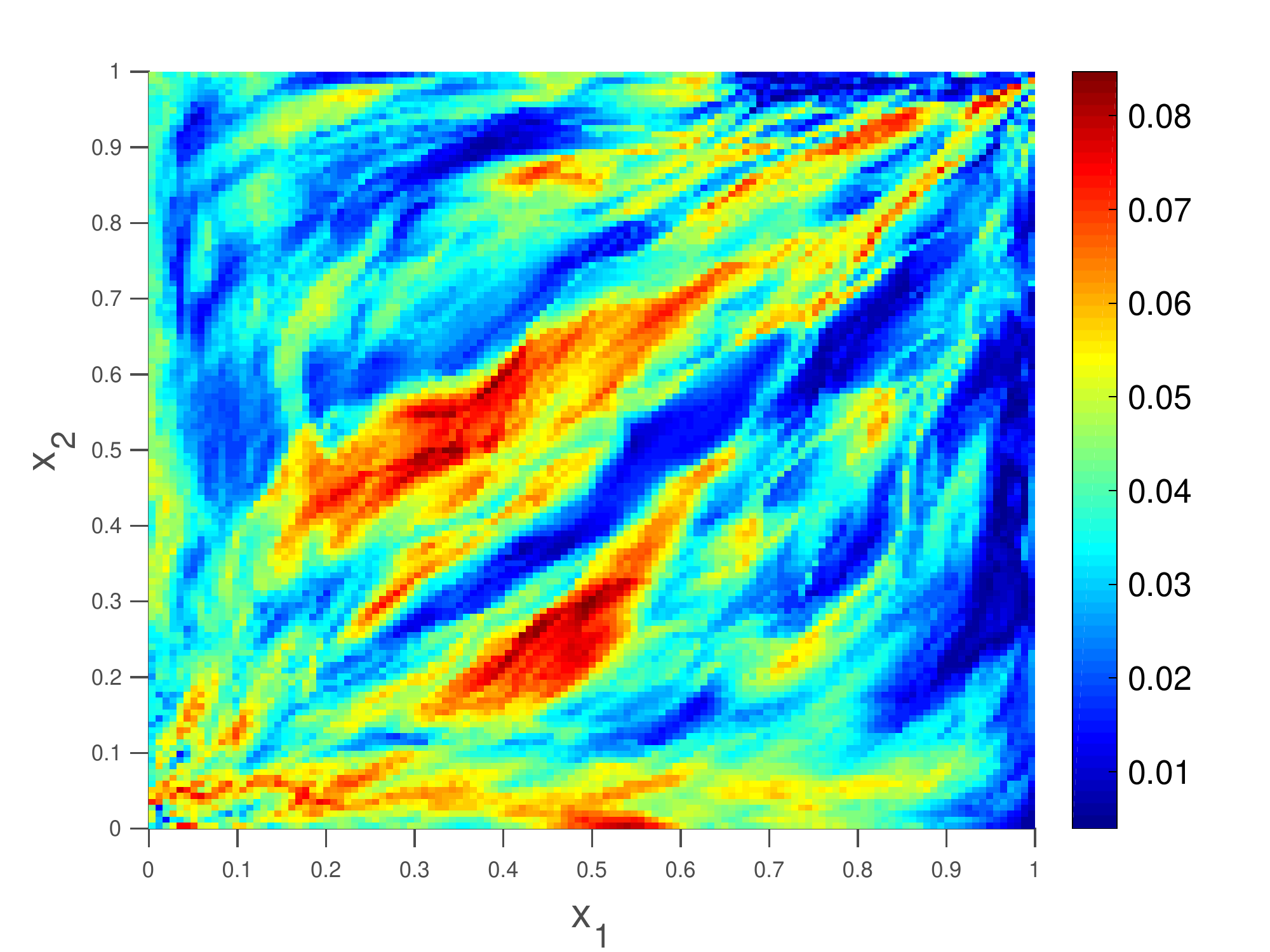}
	                \caption{Statistical distance between the $\log$(TOF) random field and its Gaussian surrogate on the entire grid. The permeability field distribution has a uniform correlation length $\lambda = 0.1L$, variance $\sigma_K^2 = 1$ and the viscosity ratio is set to be $m = 0.25$. $N_s = 2000$ realizations were used to generate the PDFs.}
	                \label{fig:TVD_logTOF}
	\end{figure}

	To provide a numerical measure of the distance between the sampled $\log$(TOF) distribution and its Gaussian surrogate, we use the statistical (TVD) distance $\delta^{\mathcal{G}}(\vec{x})$ defined as follows, 
	\begin{displaymath}
		\delta^{\mathcal{G}}(\vec{x}) = \frac{1}{2} \int_{-\infty}^{+\infty} \left| p_{\log \tau(\vec{x})}(\theta) - p_{\mathcal{G}(\log \tau(\vec{x}))}(\theta) \right| d\theta. 
	\end{displaymath}
	Intuitively, noticing that $0 \leq \delta^{\mathcal{G}}(\vec{x})  \leq 1$, $\delta^{\mathcal{G}}$ can be interpreted as the maximal probability of being able to distinguish the two random fields. \autoref{fig:TVD_logTOF} demonstrates how close both fields are and supports the use of a surrogate model for the $\log$(TOF). The maximum TVD distance is just above $0.08$, while about $80$ \% of the grid cells have a TVD distance smaller than $0.05$. 
	
	\subsection{Accuracy of the FROST distribution method}
	\label{subsection:accuracy}
	Direct MCS is the most straightforward way to estimate the saturation distribution. The method is outlined in Algorithm~\ref{alg:MCS}. Though simple, it requires a large number of realizations to be accurate, especially because the two-phase flow problem is nonlinear. Furthermore, a nonlinear numerical solver is needed to evolve the saturation in time for each realization. We consider $N_t = 4$ global time steps. For each global time step, we use an upwind scheme with $N_i = 10$ intermediate time steps to evolve the saturation along streamlines and then interpolate the saturation values back to the Eulerian grid \citep{MJM}.    
	
		\begin{algorithm}
		\begin{algorithmic}
			\For {$i = 1\cdots N_s^{MCS}$}
				\State Solve pressure equation 
				\For {$n = 1 \cdots N_t$}
					
					\For {$k = 1 \cdots N_i$}
						\State $t \gets \ds \frac{\left((n-1) + (k-1)/N_i\right)}{N_t}T$
						\State Upwind update $S_{w}^k(\vec{x}, t)^{(i)}$
						\State $S_w(\vec{x}, t)^{(i)} \gets S_{w}^k(\vec{x}, t)^{(i)}$
					\EndFor
					\State Store $S_w(\vec{x}, t)^{(i)}$
					\State Update pressure at time $t$
				\EndFor
			\EndFor
			\For {$n = 1 \cdots N_t$}
				\State $t \gets \frac{n}{N_t}T$
				\State Compute $F^{MCS}_{S_w}(s; \vec{x}, t)$ and $p^{MCS}_{S_w}(s; \vec{x}, t)$ with
				\State KDE using $\{ S_w(\vec{x}, t)^{(i)}\}_{1\leq i \leq N_s^{MCS}}$
			\EndFor
		\end{algorithmic} 
		\caption{MCS method}
		\label{alg:MCS}
		\end{algorithm}
		
\begin{figure*}
		\centering
		\includegraphics[scale=0.8]{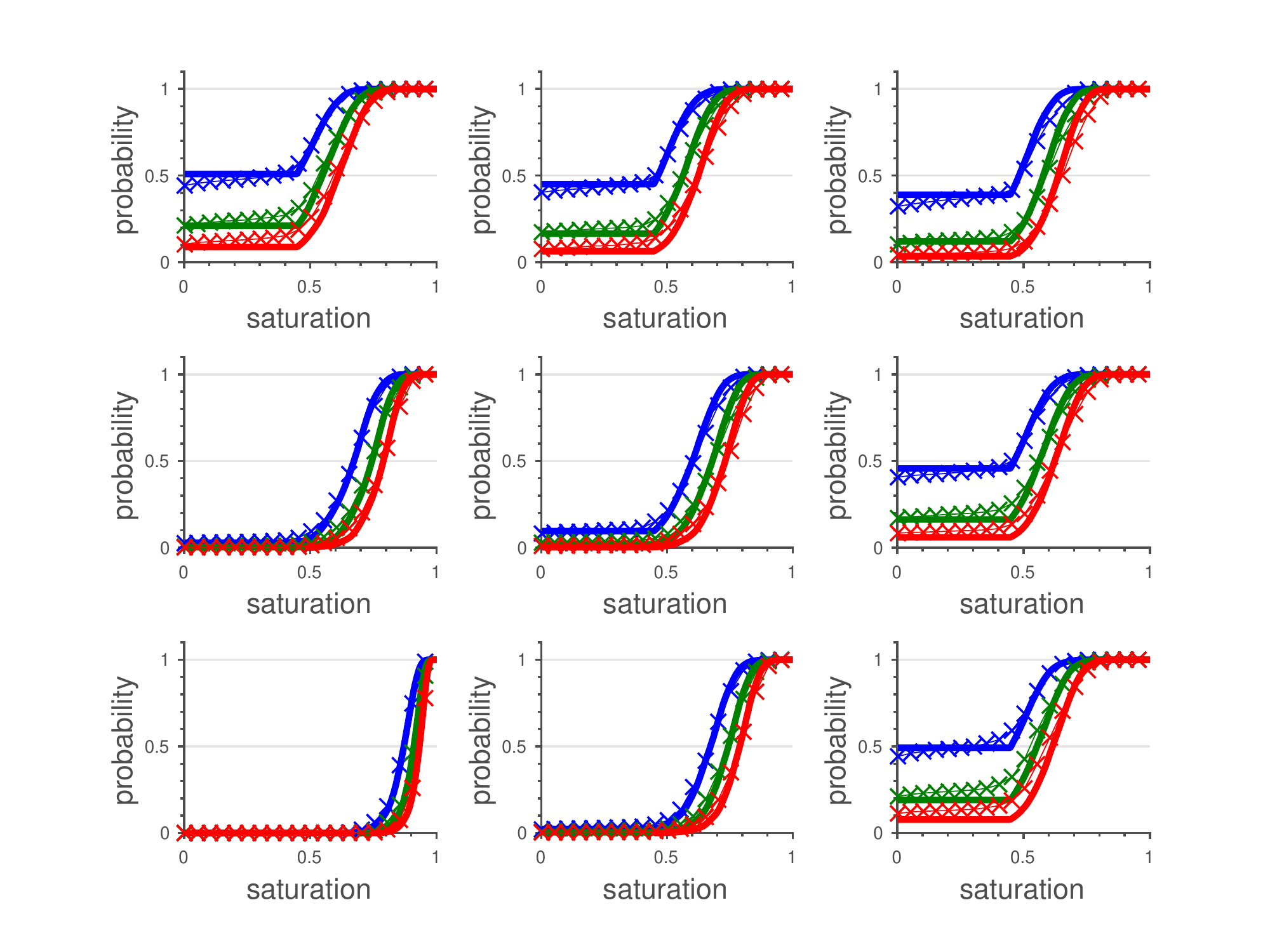}
	         \caption{Comparison between the saturation CDFs obtained with MCS (crossed lines) and FROST (solid lines) at 3 different times ($t = 0.5T$ in blue, $t = 0.75T$ in green and $t = T$ in red), and at the nine locations depicted in \autoref{fig:45spot}. The permeability field distribution has a uniform correlation length $\lambda = 0.1L$, variance $\sigma_K^2 = 1$ and the viscosity ratio is set to be $m = 0.25$. $N_s^{MCS} = N_s = 2000$ realizations were used to generate the distributions.}
	                \label{fig:cdfSwAccuracyCompare}
	\end{figure*}

\begin{figure}
		\centering
		\includegraphics[scale=0.4]{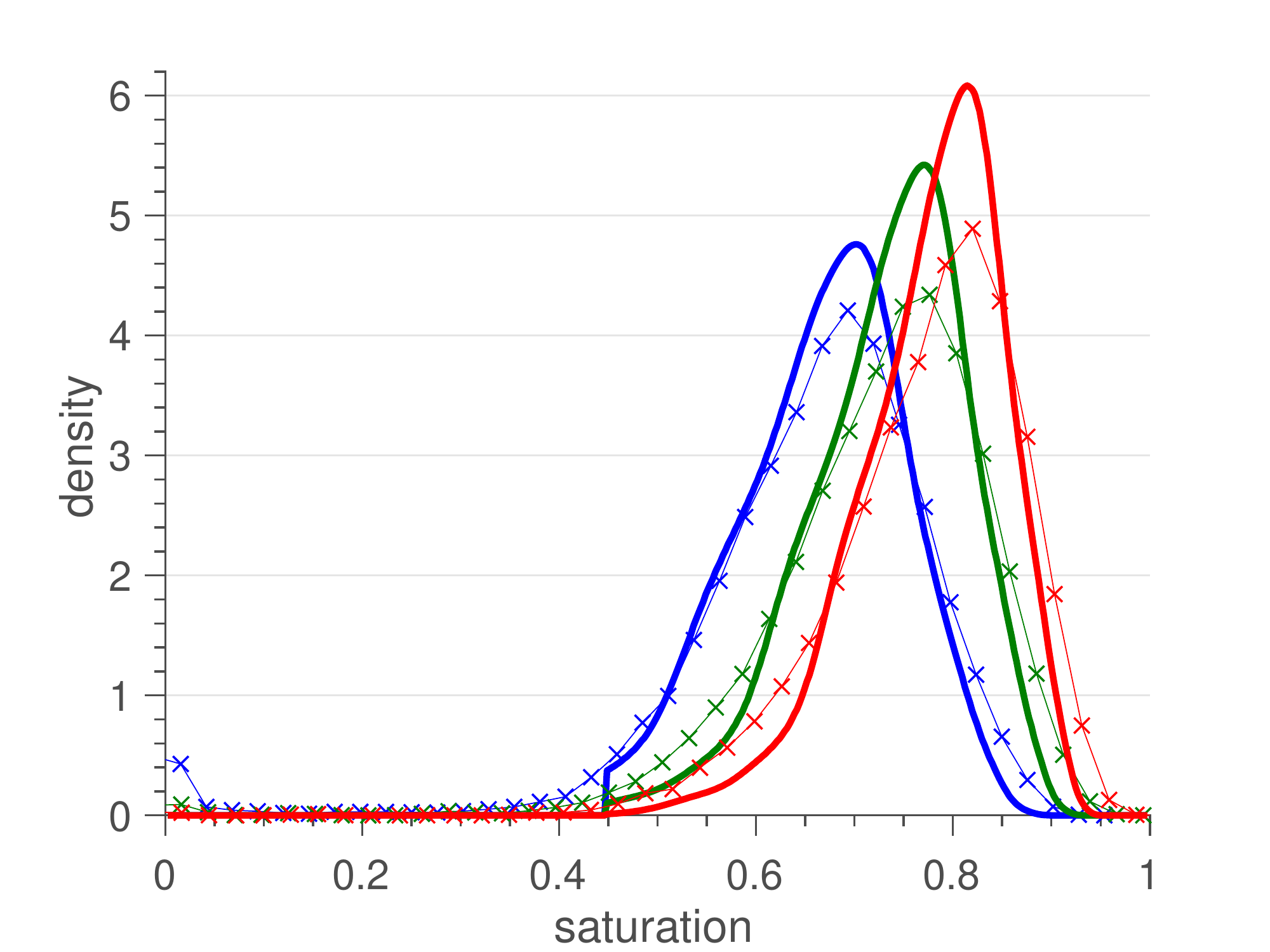}
	         \caption{Comparison between the saturation PDFs obtained with MCS (crossed lines) and FROST (solid lines) at 3 different times ($t = 0.5T$ in blue, $t = 0.75T$ in green and $t = T$ in red) and at position $(1,2)$ in \autoref{fig:45spot}. The MCS-based and FROST-based PDFs appear smooth. The permeability field distribution has a uniform correlation length $\lambda = 0.1L$, variance $\sigma_K^2 = 1$ and the viscosity ratio is set to be $m = 0.25$. $2000$ realizations were used to generate the distributions.}
	                \label{fig:pdfSwAccuracyCompare1}
	\end{figure}

\begin{figure}
		\centering
		\includegraphics[scale=0.4]{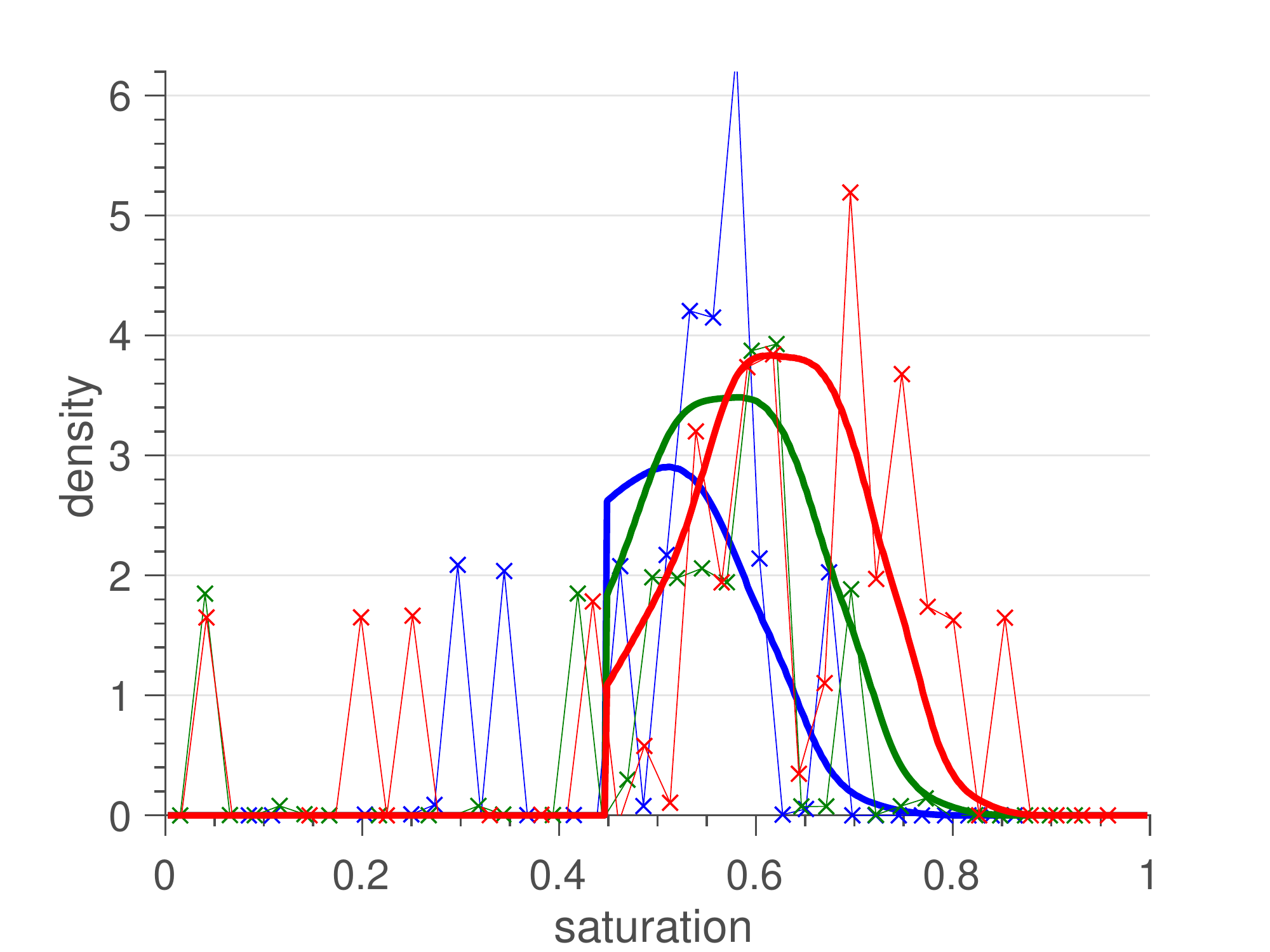}
	         \caption{Comparison between the saturation PDFs obtained with MCS (crossed lines) and FROST (solid lines) at 3 different times ($t = 0.5T$ in blue, $t = 0.75T$ in green and $t = T$ in red) and at position $(1,3)$ in \autoref{fig:45spot}. The MCS-based PDFs are oscillatory. The permeability field distribution has a uniform correlation length $\lambda = 0.1L$, variance $\sigma_K^2 = 1$ and the viscosity ratio is set to be $m = 0.25$. $N_s^{MCS} = N_s = 2000$ realizations were used to generate the distributions.}
	                \label{fig:pdfSwAccuracyCompare2}
	\end{figure}

\begin{figure*}
		\centering
		\includegraphics[scale=0.8]{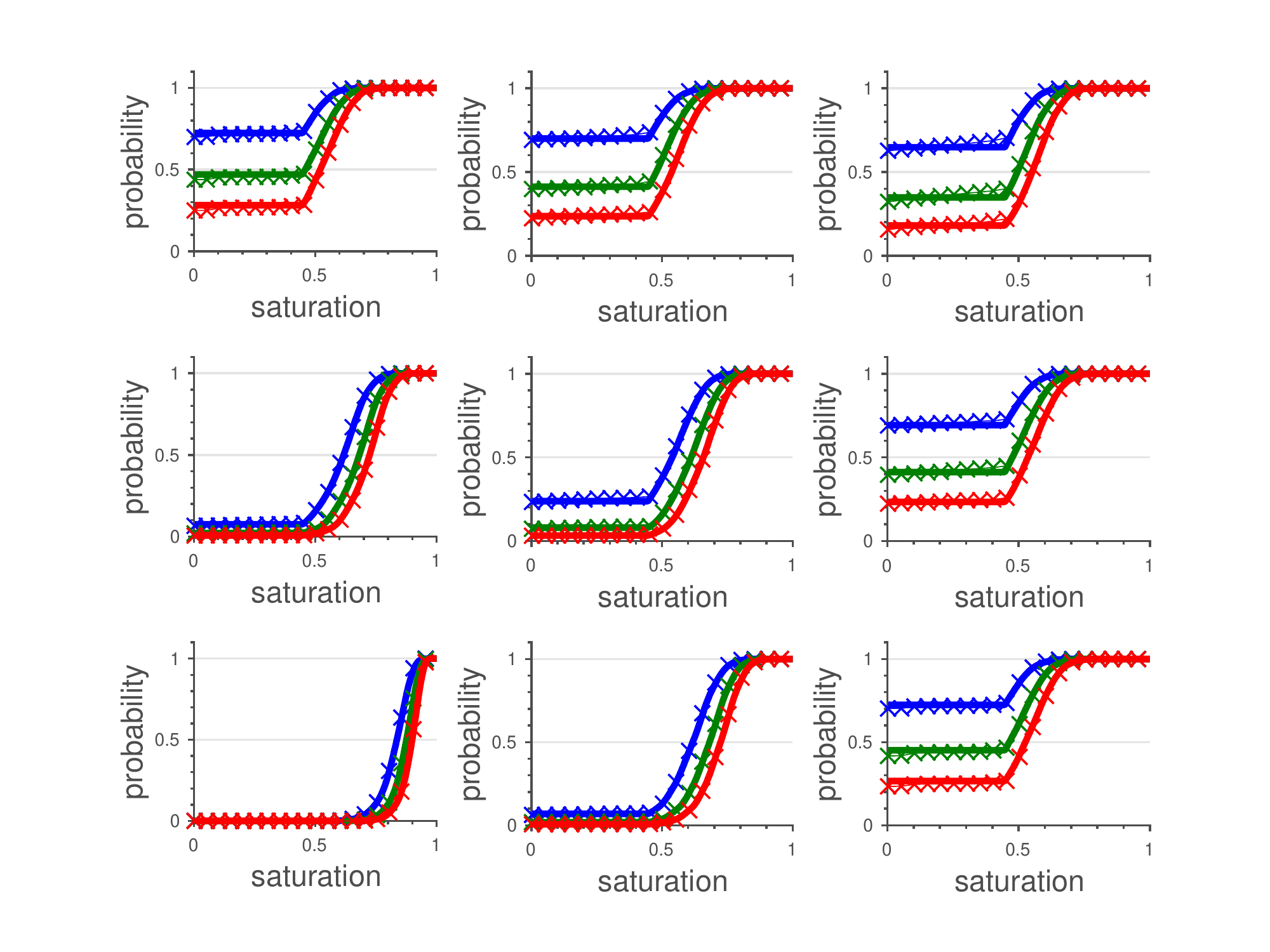}
	         \caption{Comparison between the saturation CDFs obtained with MCS (crossed lines) and FROST (solid lines) at 3 different times ($t = 0.5T$ in blue, $t = 0.75T$ in green and $t = T$ in red) and at the nine positions depicted in \autoref{fig:45spot}, with fixed flow field. The permeability field distribution has a uniform correlation length $\lambda = 0.1L$, variance $\sigma_K^2 = 1$ and the viscosity ratio is set to be $m = 0.25$. $N_s^{MCS} = N_s = 2000$ realizations were used to generate the distributions.}
	                \label{fig:cdfSwAccuracyCompareFS}
	\end{figure*}

\begin{figure}
		\centering
		\includegraphics[scale=0.4]{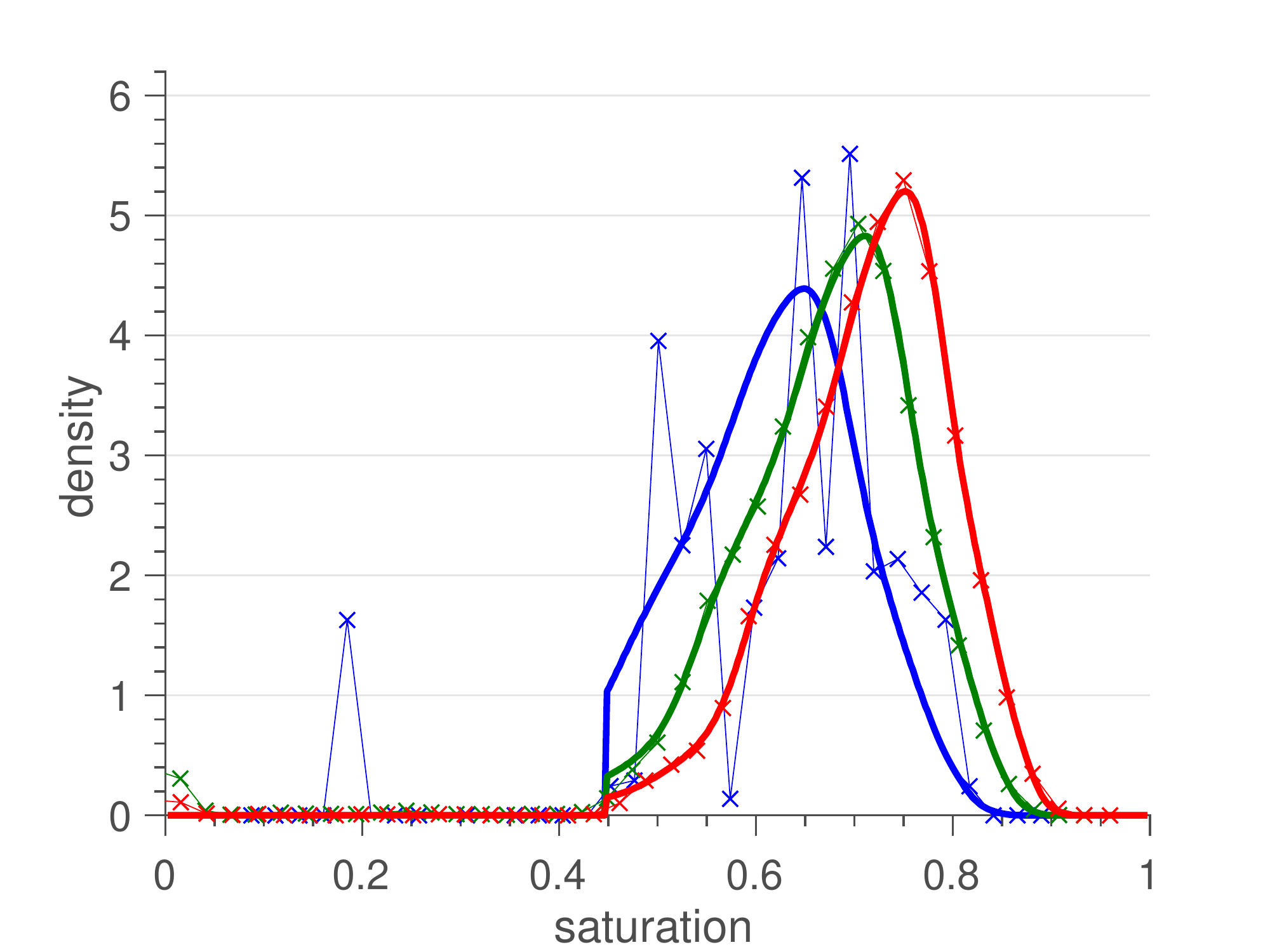}
	         \caption{Comparison between the saturation PDFs obtained with MCS (crossed lines) and FROST (solid lines) at 3 different times ($t = 0.5T$ in blue, $t = 0.75T$ in green and $t = T$ in red) with frozen streamlines, at position $(1,2)$ in \autoref{fig:45spot}. The permeability field distribution has a uniform correlation length $\lambda = 0.1L$, variance $\sigma_K^2 = 1$ and the viscosity ratio is set to be $m = 0.25$. $N_s^{MCS} = N_s = 2000$ realizations were used to generate the distributions.}
	                \label{fig:pdfSwAccuracyCompareFS1}
	\end{figure}

\begin{figure}
		\centering
		\includegraphics[scale=0.4]{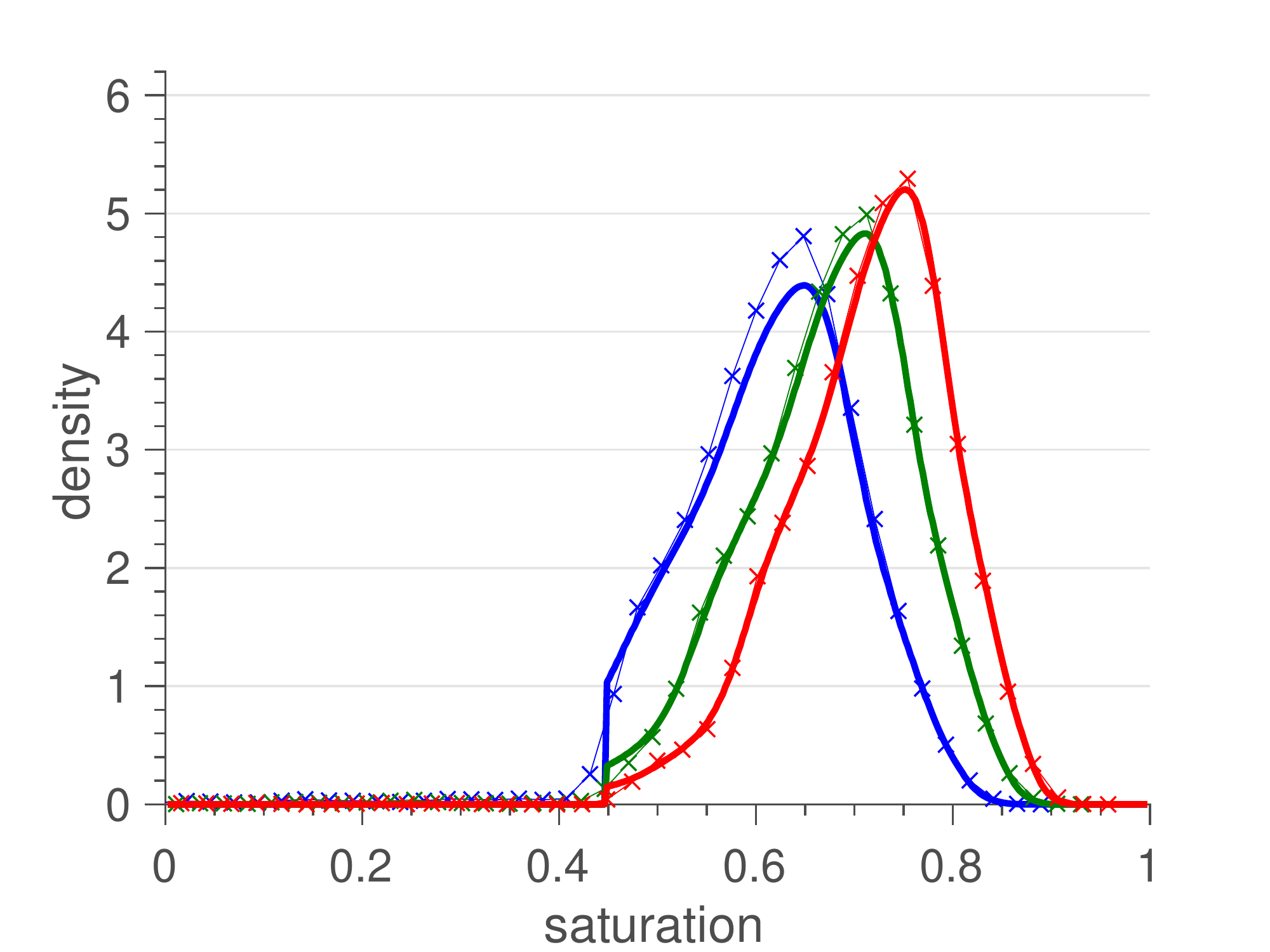}
	         \caption{Comparison between the saturation PDFs obtained with ``smoothed" MCS (crossed lines) and FROST (solid lines) at 3 different times ($t = 0.5T$ in blue, $t = 0.75T$ in green and $t = T$ in red) with frozen streamlines, at position $(1,2)$ in \autoref{fig:45spot}. The permeability field distribution has a uniform correlation length $\lambda = 0.1L$, variance $\sigma_K^2 = 1$ and the viscosity ratio is set to be $m = 0.25$. When smoothed, the PDF is adequately captured. $N_s^{MCS} = N_s = 2000$ realizations were used to generate the distributions.}
	                \label{fig:pdfSwAccuracyCompareFS}
	\end{figure}

We compare the CDF and PDF of the saturation obtained using direct MCS on the saturation realizations versus those obtained by performing the FROST with $\log$(TOF) distribution estimated using $\log$(TOF) realizations and KDE, and the mean EIT coming from the power law fit with $M_s = 100$, further analyzed in \autoref{subsection:EIT}. \autoref{fig:cdfSwAccuracyCompare} compares the saturation CDFs using the FROST and MCS, with $m = 0.25$, $\sigma^2_K = 1$, $\lambda = 0.1L$ and $N_s = 2000$ and for three different times. We can see that the FROST is in good agreement with MCS. For positions close to the injection, the CDF is smooth and we observe slight discrepancies between FROST and MCS. Close to the producer, the FROST predicts that the CDFs have a plateau. These plateaus seem to be present in the MCS-based CDFs, even though they are slightly smeared out. This is mainly due to our use of a basic streamline solver for MCS that results in smearing effects when mapping saturations between the Eulerian flow-problem grid and streamlines \citep{MG,MGM}. 
While there is reasonable agreement in \autoref{fig:pdfSwAccuracyCompare1} close to the injector, in \autoref{fig:pdfSwAccuracyCompare2} we show that KDE can fail to estimate the saturation PDF when it is not smooth. The MCS-based PDFs, close to the producer, are oscillatory. This is expected since the PDF is a combination of a Dirac and a continuous regions (see \autoref{eq:multiDPDFsaturation}) and KDE does not necessarily converge for non-continuous distributions. (The saturation PDF is not continuous since the CDF is only piecewise $\mathcal{C}^{1}$ on $(0,1)$ and can have a jump at $s=0$.) To verify that the FROST outputs a coherent estimate, we compare the FROST with MCS when we indeed fix the entire flow field (hence the streamlines are fixed and the streamtube capacities are not updated in time) and use a large number of streamlines and streamline points to minimize the numerical effect in the MCS. In this fixed flow field setting, $\langle EIT \rangle (t) = t$ exactly. \autoref{fig:cdfSwAccuracyCompareFS} shows that the FROST-based CDF indeed correspond to the MCS-based CDF. In \autoref{fig:pdfSwAccuracyCompareFS1}, we illustrate that the MCS-based PDF estimate may not converge, while in \autoref{fig:pdfSwAccuracyCompareFS}, we display a ``smoothed" version of these MCS-based PDFs obtained by removing realizations where the saturation is zero, hence avoiding the non-smooth behavior of the distribution at $s=0$. The displayed ``smoothed" MCS-based PDF is in agreement with the shape of the estimated FROST based PDFs, while, as expected, overestimating the peak of the density where the PDF is non-smooth.  


	\subsection{Efficiency of sampling and accelerated FROST method}
	\label{subsection:efficiency}
	
\begin{figure}
			\centering
			\includegraphics[scale=0.4]{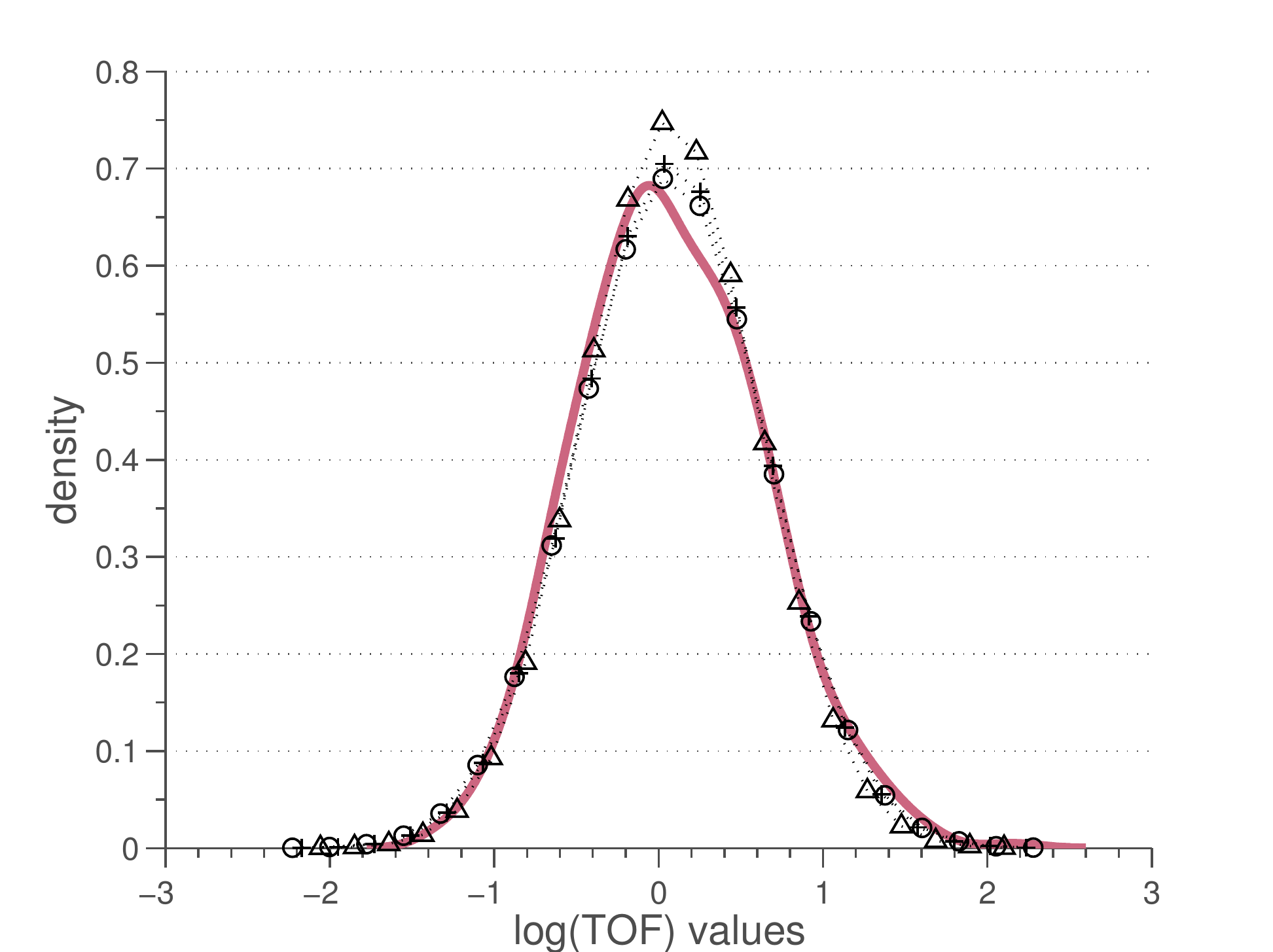}
	        		 \caption{Comparison of the $\log$TOF PDFs, at location $(2,2)$ in \autoref{fig:45spot}, using samples directly (solid lines) and surrogate Gaussian distributions using 80 (triangles), 400 (circles) and 2000 (pluses) realizations. The permeability field distribution has a uniform correlation length $\lambda = 0.1L$, variance $\sigma_K^2 = 4$ and the viscosity ratio is set to be $m = 0.25$.}
	               	 \label{fig:effSampling}
		\end{figure}

\begin{figure*}
			\centering
			\includegraphics[scale=0.8]{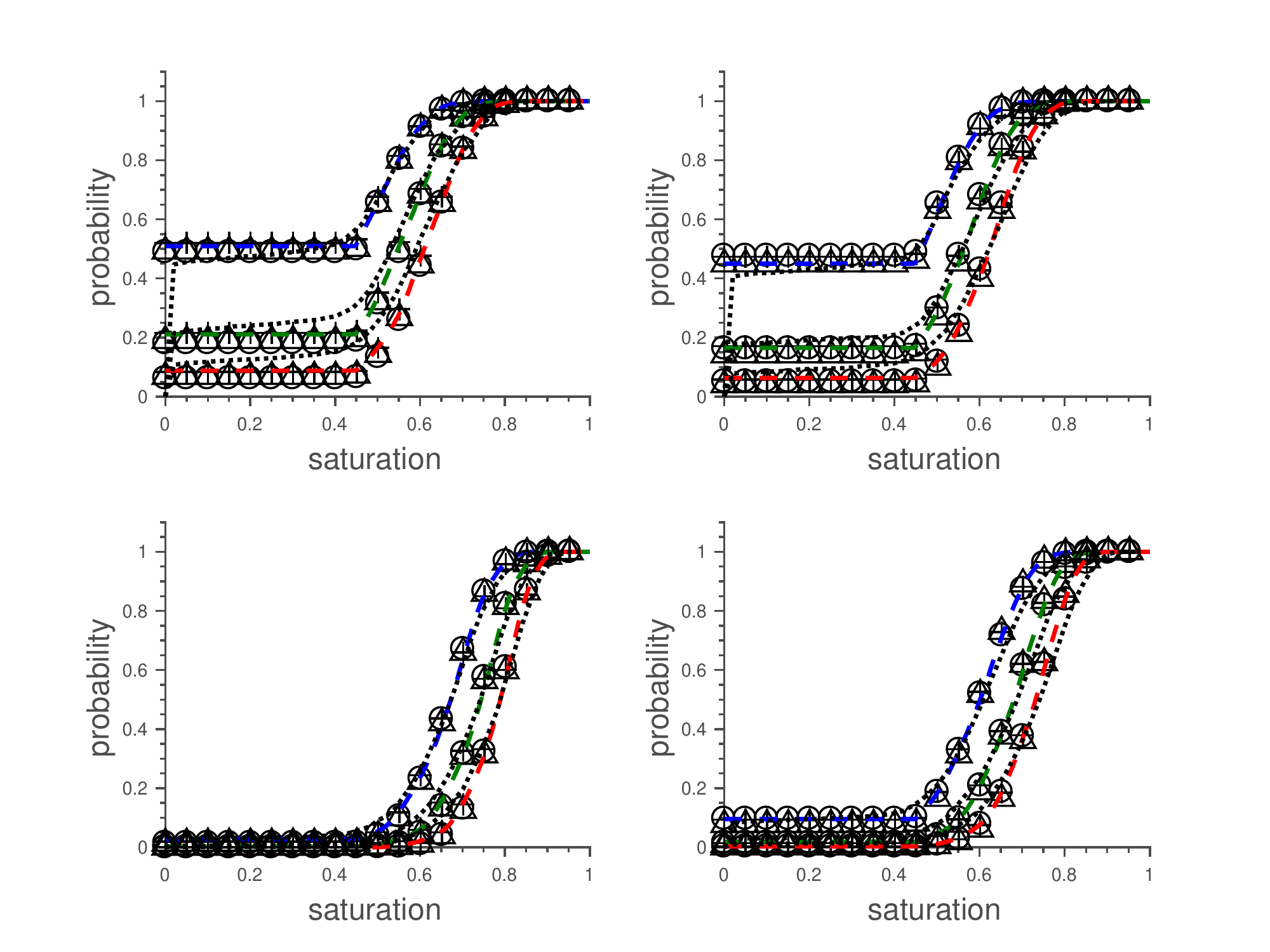}
	        		 \caption{Comparison of the saturation CDFs using MCS (black dashed) with $N_s^{MCS} = 2000$, FROST (colored dashed) with $N_s = 2000$, and gFROST using $N_s^g = 80$ (triangles), $N_s^g = 400$ (circles) and $N_s^g = 2000$ (pluses) realizations at positions $(2,3)$ (top left), $(3,3)$ (top right), $(2,2)$ (bottom left) and $(3,2)$ (bottom right) from \autoref{fig:45spot}. The permeability field distribution has a uniform correlation length $\lambda = 0.1L$, variance $\sigma_K^2 = 1$ and the viscosity ratio is set to be $m = 0.25$.}
	               	 \label{fig:cdfSwCompareSurr5krExpL01V05M14}
		\end{figure*}

		We previously argued that the $\log$(TOF) distributions could be well approximated by Gaussian distributions (see \autoref{subsection:Simplific}, \autoref{subsection:smoothness}, \autoref{fig:logTOF_pdfs1}, and \autoref{fig:TVD_logTOF}). \autoref{fig:effSampling} shows that a reliable estimate of the $\log$(TOF) distribution can be achieved already with few realizations (in the order of $10$ times less as compared to KDE) for the mean and variance sample estimates. This leads to further computational cost savings of the FROST method. \autoref{fig:cdfSwCompareSurr5krExpL01V05M14} demonstrates the accuracy of the Gaussian surrogate FROST versus direct MCS and FROST for the saturation distribution. This means that we can preprocess the distribution of the $\log$(TOF) by solving a few steady-state problems, and reuse this distribution to estimate the saturation distribution for any desired time. We can therefore formulate a fast and efficient algorithm to estimate the saturation distribution in Algorithm \ref{alg:GsDM}, where $N^g_s$, the number of realizations needed, is typically smaller than $N_s$, the number of realizations used in the FROST method, and $M_s$ is the same as in the FROST method. We will refer to this method as gFROST.
		\begin{algorithm}
		\begin{algorithmic}
			\For {$i = 1\cdots N^g_s$}
				\State Solve initial tracer equation 
				\State Store time-of-flights $\tau(\vec{x})^{(i)}$
                \If {$i \leq M_s$}
                	\State Solve transport eq. at times $T/N_t$ and $2T/N_t$
                	\State Store velocity fields $\left\{q^{(i)}_{tot}(kT/N_t)\right\}_{k=1,2}$
                 \EndIf
			\EndFor
			\State Compute sample mean, $m^{(N_s^{g})}(\vec{x})$, and standard deviation, $s^{(N_s^{g})}(\vec{x})$ of $\left\{ \tau(\vec{x})^{(i)} \right\}_{1 \leq i \leq N^{g}_s}$
			\State Estimate $p_{\log\tau(\vec{x})}$ and $F_{\log\tau(\vec{x})}$ with $\log\tau(\vec{x}) \sim \mathcal{N}[m^{(N_s^{g})}(\vec{x}), (s^{(N_s^{g})}(\vec{x}))^2]$
			\State Estimate $c$, $\beta$ and $\langle EIT \rangle (t) = c t^\beta$ using $\left\{q^{(i)}_{tot}(kT/N_t)\right\}_{k=1,2; 1\leq i \leq N_s}$
			\State Compute $\left\{F^{gFROST}_{S_w}(s; \vec{x},t_j)\right\}_{j = 1 \dots N_t}$ and $\left\{p^{gFROST}_{S_w}(s; \vec{x}, t_j)\right\}_{j = 1 \dots N_t}$ for $t_j = j T/N_t$ 
		\end{algorithmic} 
		\caption{Gaussian surrogate based FROST (gFROST)}
		\label{alg:GsDM}
		\end{algorithm}
	
\subsection{Study of EIT behavior}
\label{subsection:EIT}

\begin{figure*}
		\centering
			 \includegraphics[width=\textwidth]{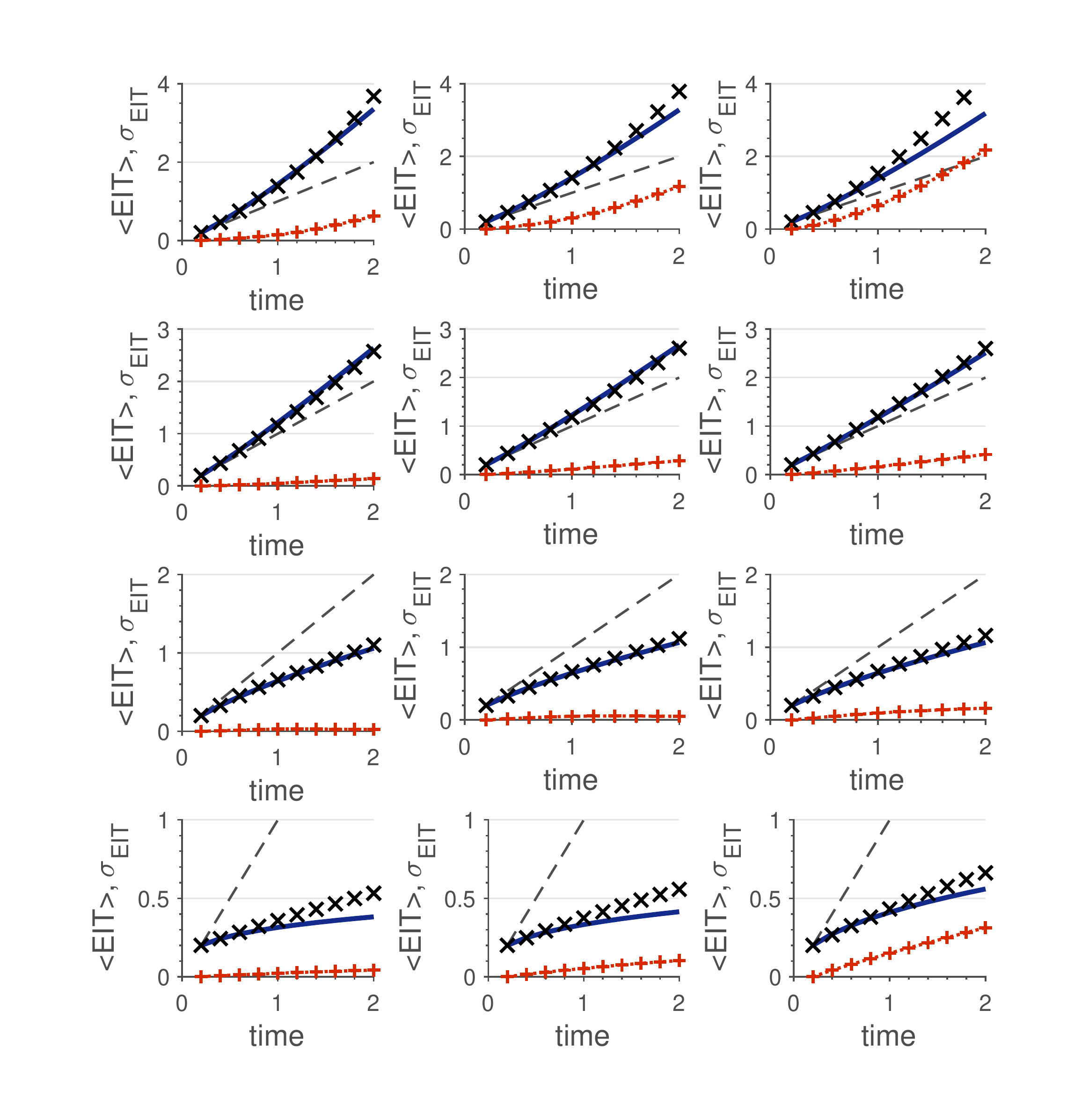}
	         \caption{Plots of $\langle EIT \rangle$ (crosses) and $\sigma_{EIT}$ (pluses) over time for pressure control. Row 1, 2, 3 and 4 respectively correspond to $m=0.1$, $m=0.5$, $m=2$ and $m=10$, while column 1, 2 and 3 respectively correspond to $\sigma_K^2=0.25$, $\sigma_K^2=1$ and $\sigma_K^2=4$. The solid line corresponds to the fitted power law EIT and the dashed line indicates the slope $\langle EIT \rangle (\Delta t)/\Delta t$ for reference. $M_s = 200$ realizations of the $EIT$ were used.}
	                \label{fig:EITsensitivity}
	\end{figure*}

\begin{figure*}
		\centering
			 \includegraphics[width=\textwidth]{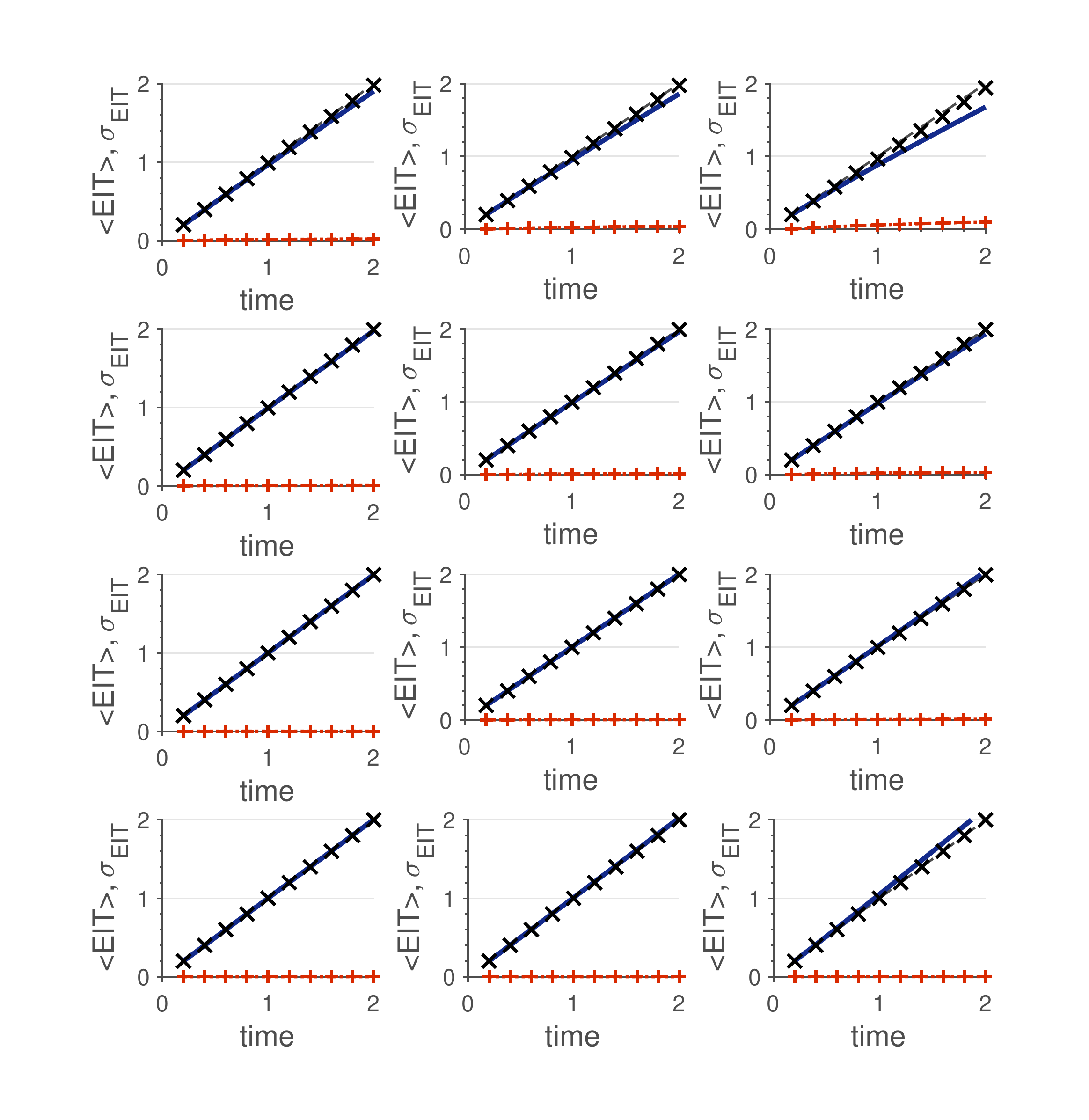}
	         \caption{Plots of $\langle EIT \rangle$ (crosses) and $\sigma_{EIT}$ (pluses) over time for rate control ($\mathbf{q}_{inj} = \mathbf{e}_{y}$ and $\mathbf{q}_{prod} = \mathbf{0}$). Row 1, 2, 3 and 4 respectively correspond to $m=0.1$, $m=0.5$, $m=2$ and $m=10$, while column 1, 2 and 3 respectively correspond to $\sigma_K^2=0.25$, $\sigma_K^2=1$ and $\sigma_K^2=4$. The solid line corresponds to the fitted power law EIT and the dashed line indicates the slope $\langle EIT \rangle (\Delta t)/\Delta t$ for reference. $M_s = 200$ realizations of the $EIT$ were used.}
	                \label{fig:EITsensitivityR}
	\end{figure*}

\begin{figure}
		\centering
		\includegraphics[scale=0.4]{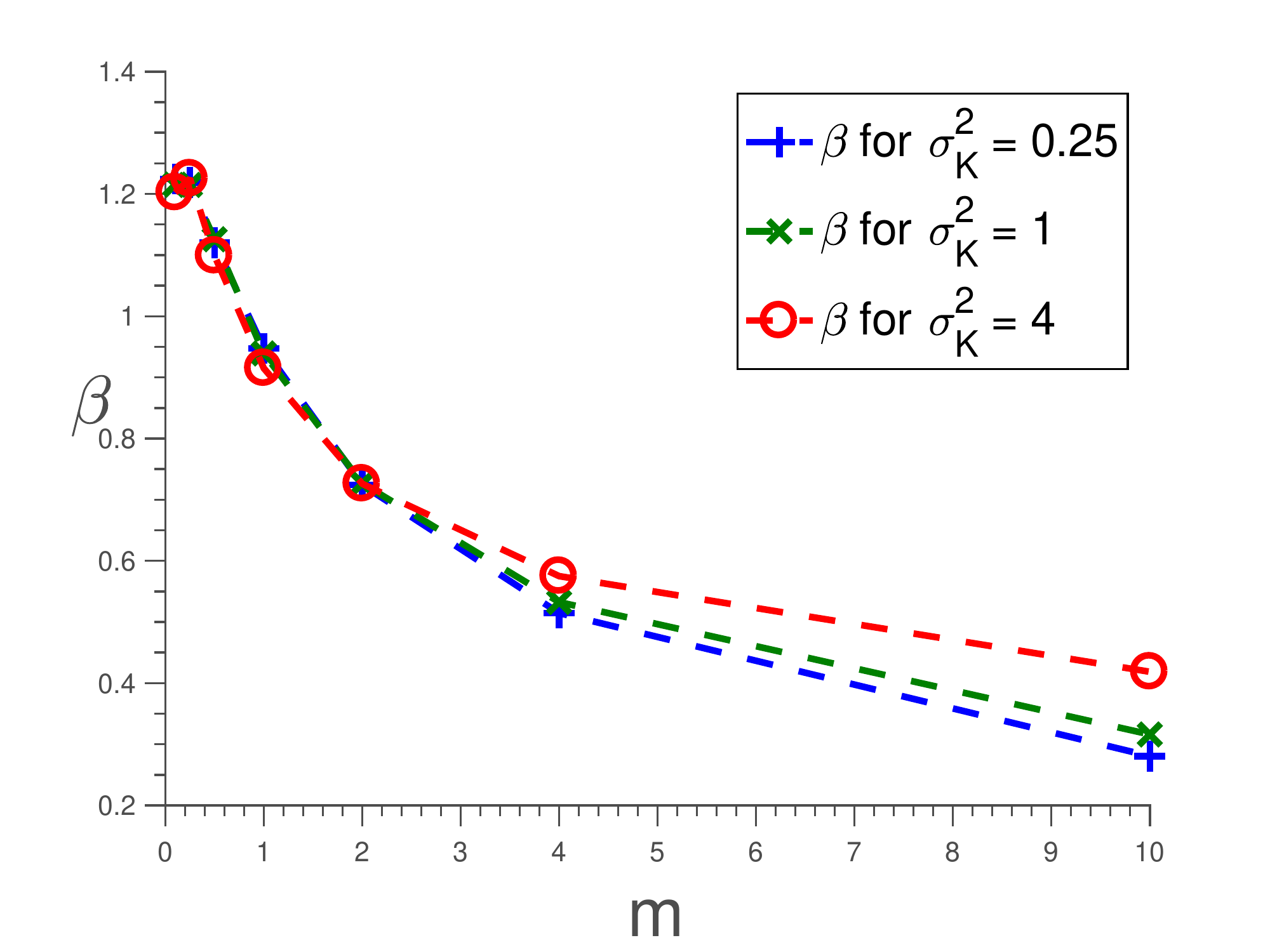}
	         \caption{Power law fitting exponent $\beta$ of $\langle EIT \rangle (t)$ as a function of the viscosity ratio $m$.}
	                \label{fig:beta_fitting}
	\end{figure}

\begin{figure}
		\centering
		\includegraphics[scale=0.4]{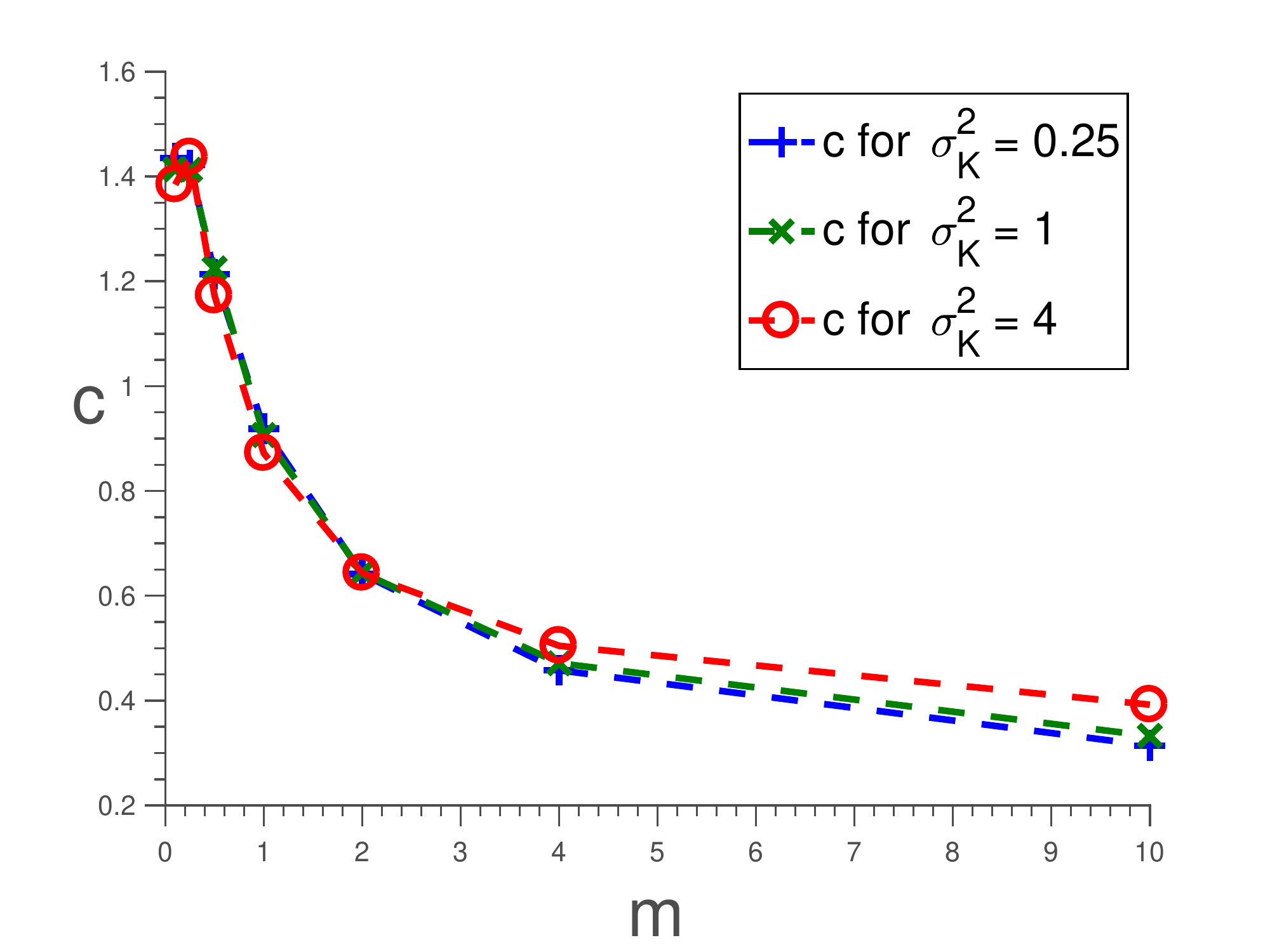}
	         \caption{Power law fitting intercept $c$ of $\langle EIT \rangle (t)$ as a function of the viscosity ratio $m$.}
	                \label{fig:c_fitting}
	\end{figure}

In \autoref{section : methodology}, we argued that the EIT was almost deterministic (and hence could be estimated by its average $\langle EIT \rangle$ with a few samples) and that it approximately followed a power law in time. We study here when this is indeed the case by investigating the sensitivity of the EIT with respect to the viscosity ratio $m$ and the permeability field variance $\sigma_K^2$. 
We assume that the permeability field has a log normal distribution with exponential covariance and $\sigma_K^2 \in \{ 0.25, \; 1, \; 4 \}$, $\lambda = 0.1L$. We use $M_s = 200$ flow realizations with pressure updates. We vary the viscosity ratio to be $m = \mu_w / \mu_o \in \{ 0.1, \; 0.25, \; 0.5, \; 1, \; 2, \; 4, \; 10 \}$. \autoref{fig:EITsensitivity} shows that, under pressure control, the mean EIT is following a power law to a good approximation. On this plot, we display the mean EIT and its standard deviation at position $(2,2)$ only since, remarkably, these two statistics do not depend on the position $\vec{x}$ in the reservoir. This spatial independence is another advantage of the FROST. However, the EIT is not always guaranteed to have a small variance. The EIT standard deviation is negligible for viscosity ratios $m$ close to 1 (either slightly  smaller or larger than 1). Hence, this regime corresponds to the domain of applicability of the FROST with the approximation $EIT \approx \langle EIT\rangle$. Indeed, for viscosity ratios close to unity, the EIT variance remains small even for large input variance $\sigma_K^2$ and the power law fitting is satisfactory. In addition, \autoref{fig:beta_fitting} and \autoref{fig:c_fitting} show that the power law fitting parameters for the mean EIT, $\beta$ and $c$, not only decay exponentially with increasing $m$ but are also weakly sensitive to the input variance $\sigma_K^2$, at least for $m \leq 4$. This means that we can efficiently fit $c(m)$ and $\beta(m)$, and only sample for the TOF in the FROST to be able to estimate the saturation distribution.
Moreover, for rate control, \autoref{fig:EITsensitivityR} shows that the EIT is in this case almost deterministic regardless of the level of $\sigma_{K}^2$ and its mean is linear, rather than following a power law. The non-zero EIT variance for $m<1$ is most probably resulting from the unstable displacement of more viscous oil by less viscous water that induces streamline-pattern changes and EIT variability. 
	
In conclusion, in the pressure control case, within the frozen streamline approximation, the validity of the reduction of the EIT distribution to $\langle EIT \rangle$ is accurate for $m$ of the order of $1$. While for $m = 0.5 \dots 2$ the EIT standard deviation is relatively small, for $m=0.1$ and $10$ and $\sigma_K^2 = 4$, the standard deviations become comparable to $\langle EIT \rangle$. Hence, the (simplified) FROST is accurate for viscosity ratios close to unity, and the mean EIT can be preprocessed thanks to its power law parametrization. Nonetheless, for small and large viscosity ratios, we can still use the simplified FROST but lose accuracy in the saturation distribution estimation or work directly with the full EIT distribution. In the rate control case, the EIT approximation is almost exact. 

\FloatBarrier
	
\section{Using FROST with geologically realistic porous systems}
\label{section : numres}

		\subsection{Setting: a highly anisotropic porous medium}
		\label{subsection : two-phase}
			
\begin{figure}
		\centering
		\includegraphics[scale=0.4]{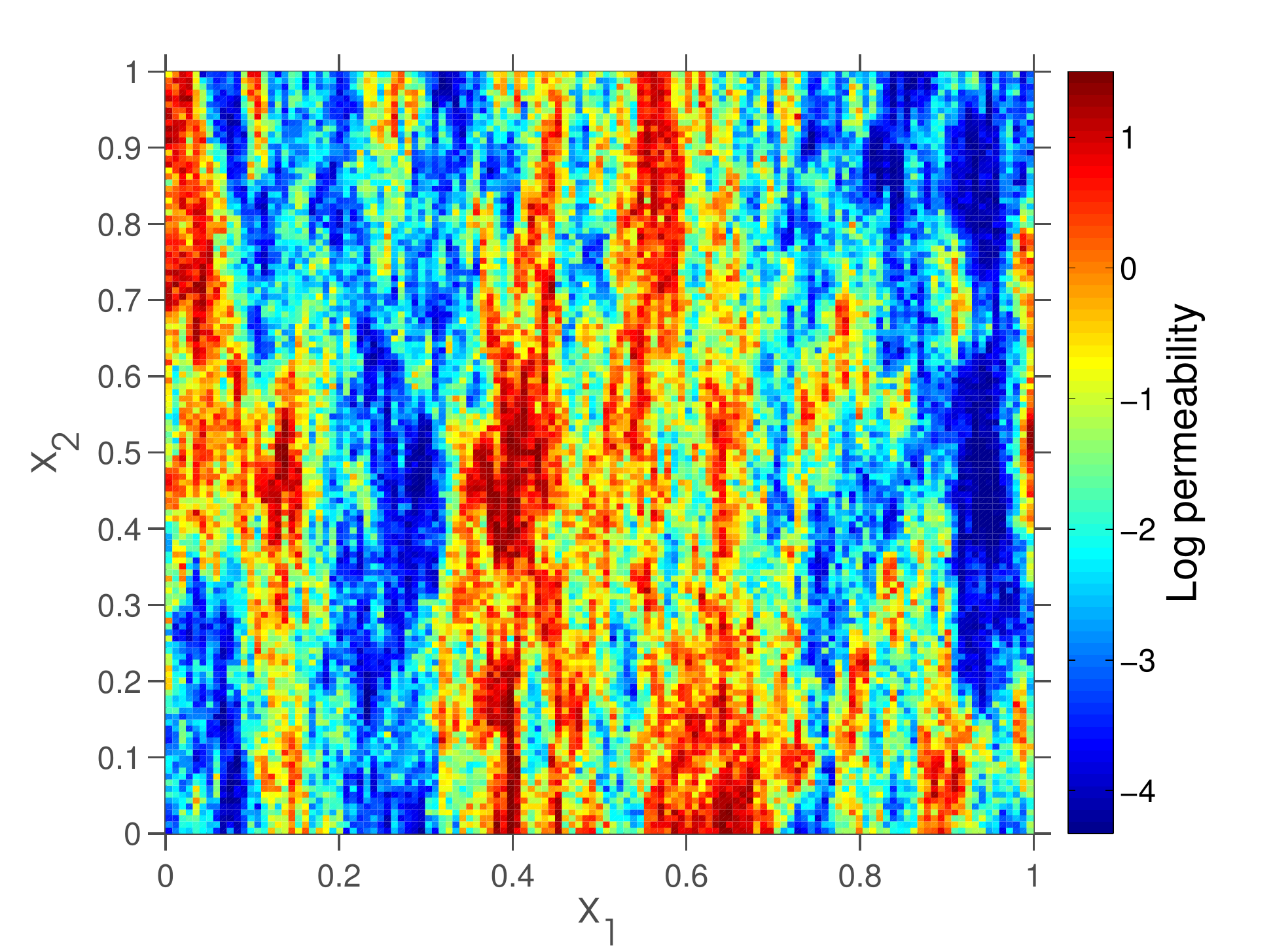}
	         \caption{A realization of anisotropic layered permeability field generated by variogram-based geostatistics.}
	                \label{fig:realOF}
	\end{figure}

		We illustrate how the FROST and gFROST methods, with EIT approximated by the power law in \autoref{eq:meanEIT_power}, can be used to estimate saturation statistics. We provide qualitative comparisons of the FROST, gFROST and MCS when the permeability field is assumed to be highly heterogeneous due to anisotropy. We consider the same setting as in \autoref{section : companal}, i.e. the quarter-five spot configuration with pressure control, but the permeability field is generated by a variogram-based geostatistical model instead of log-normal distribution. This is done by using the Geostatistical Earth Modeling Software (GEMS), first introduced in \cite{DJ}. We generate $3000$ realizations of (anisotropic) oriented features for the permeability field with exponential variogram in the $x_1$ direction, 
		\begin{displaymath}
			\gamma_{1}(\vec{h}) = c_1 \left[ 1 - \exp\left( - \frac{3 \left\|\vec{h}\right\|}{a_1}\right)\right],
		\end{displaymath}
and analogously in the $x_2$ direction, with sills $c_1 = c_2 = 1$, and practical ranges $a_1 = 10 L/128 \approx 0.08 L$ and $a_2 = 200 L/128 \approx 1.56 L$. The distribution of permeabilities has variance $\sigma_K^2 = 1.5$ (a typical realization is presented in \autoref{fig:realOF}). Besides comparing the saturation mean and standard deviation between the FROST, gFROST and MCS, we introduce the notion of quantiles and likely scenarios for the saturation field.

	\subsection{Results}
	\label{subsection : results}
		
		\subsubsection{Saturation CDF}

\begin{figure*}
		\centering
		\includegraphics[scale=0.8]{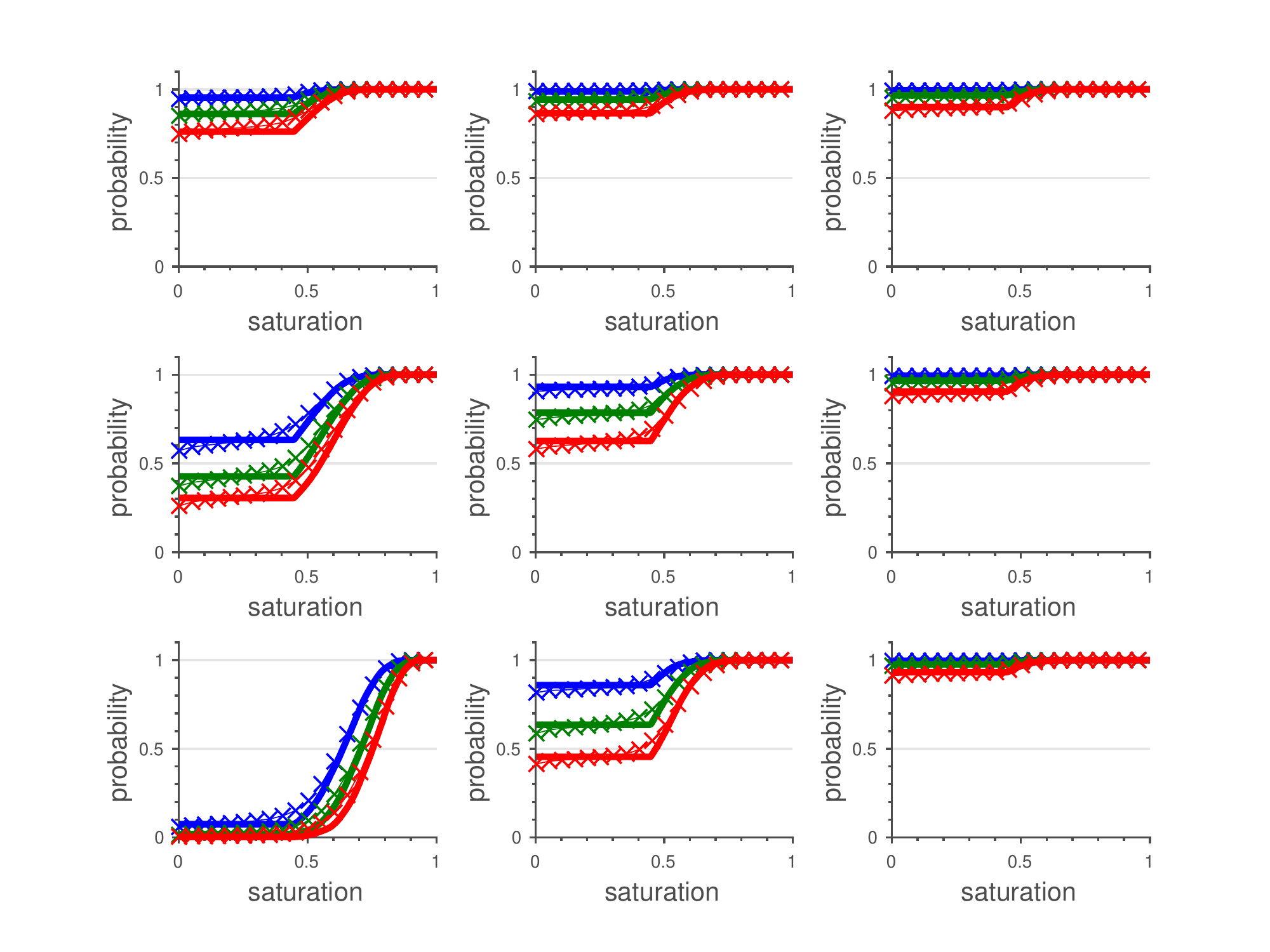}
	         \caption{Comparison between the saturation CDFs obtained with MCS (crossed lines) and FROST (solid lines) at 3 different times ($t = 0.5T$ in blue, $t = 0.75T$ in green and $t = T$ in red) and at the nine positions from \autoref{fig:45spot}.  The stochastic permeability field is highly anisotropic in the $x_2$ direction. The viscosity ratio is set to be $m = 0.25$. $N_s^{MCS} = N_s = 3000$ realizations were used to generate the distributions.}
	                \label{fig:cdfSwOFCompare}
	\end{figure*}

\autoref{fig:cdfSwOFCompare} displays the FROST-based and MCS-based saturation CDFs at different times for the numerical test. We use here $N_s = N_s^{MCS} = 3000$ realizations for both FROST and MCS estimates. $M_s = 100$ two-time steps solves of transport were performed to estimate $\langle EIT \rangle$. We can see that, as explained in \autoref{subsection:accuracy}, the MCS results tend to smoothen plateaus. But overall, the trends are captured.  
	
		\subsubsection{Mean and standard deviation of saturation}
			
			In \autoref{section : companal}, we discussed how the FROST and gFROST performed against MCS in estimating the saturation PDF and CDF. Even though estimating the saturation distributions is the main purpose of this paper, we showcase results for the first (mean) and second (standard deviation) moments of the saturation using the three approaches. These two statistics are broadly used in the literature as uncertainty quantification tools and provide convenient information to construct rapid confidence intervals of the saturation variability. \autoref{eq:multiDCDFm1} enables the computation of any saturation moment. In particular, for the mean and standard deviation we have 
\begin{displaymath}
	\langle S_w \rangle^{}(\vec{x},t) = 1 - \int_0^1 F^{}_{S_w}(s; \vec{x},t) ds 
\end{displaymath}
and
\begin{displaymath}
	\sigma_{S_w}^{}(\vec{x},t)^2 = 1 - \int_0^1 F^{}_{S_w}(\sqrt{s}; \vec{x},t) ds - \langle S_w \rangle^{}(\vec{x},t)^2.
\end{displaymath}
			
\begin{figure*}
		\centering
		\includegraphics[width=\textwidth]{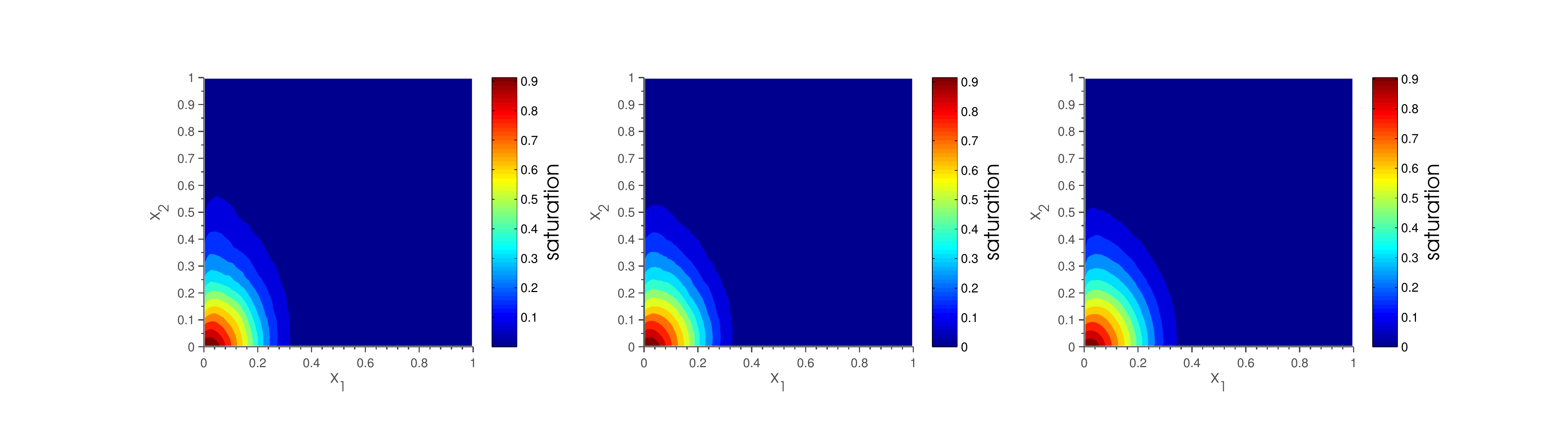}
	         \caption{Comparison between the mean saturation obtained with gFROST (left), FROST (middle) and MCS (right) at time $t = 0.25T$. The stochastic permeability field is highly anisotropic in the $x_2$ direction. The viscosity ratio is set to be $m = 0.25$. $N_s^{MCS} = 3000$ realizations were used for MCS,  $N_s = 1000$ for FROST and $N_s^{g} = 300$ for gFROST.}
	                \label{fig:meanSwComparet1}
	\end{figure*}

\begin{figure*}
		\centering
		\includegraphics[width=\textwidth]{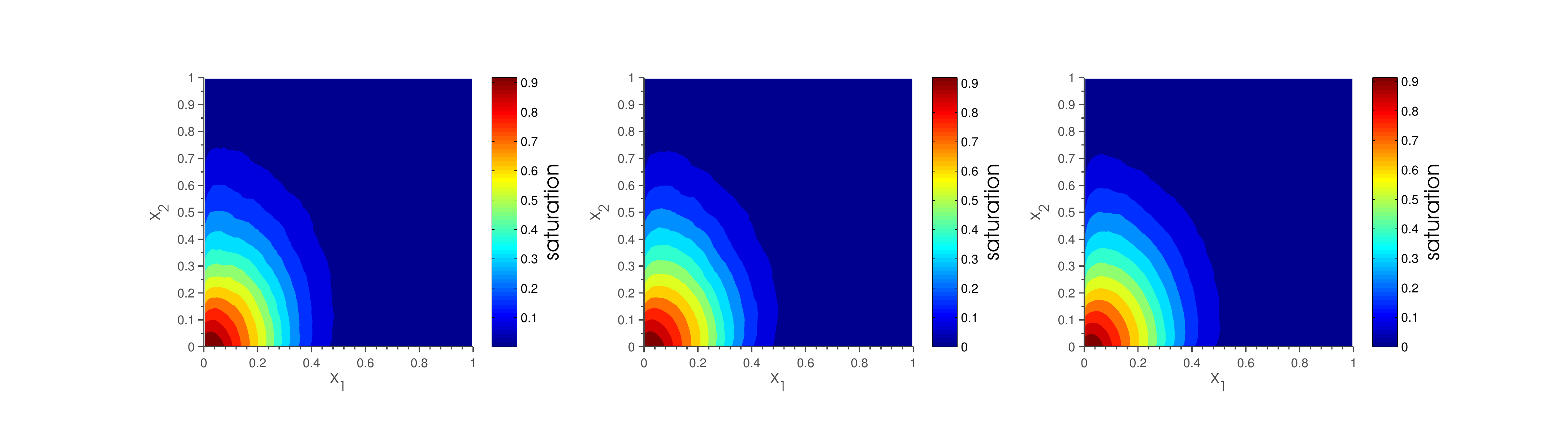}
	         \caption{Comparison between the mean saturation obtained with gFROST (left), FROST (middle) and MCS (right) at time $t = 0.5T$. The stochastic permeability field is highly anisotropic in the $x_2$ direction. The viscosity ratio is set to be $m = 0.25$. $N_s^{MCS} = 3000$ realizations were used for MCS,  $N_s = 1000$ for FROST and $N_s^{g} = 300$ for gFROST.}
	                \label{fig:meanSwComparet2}
	\end{figure*}

\begin{figure*}
		\centering
		\includegraphics[width=\textwidth]{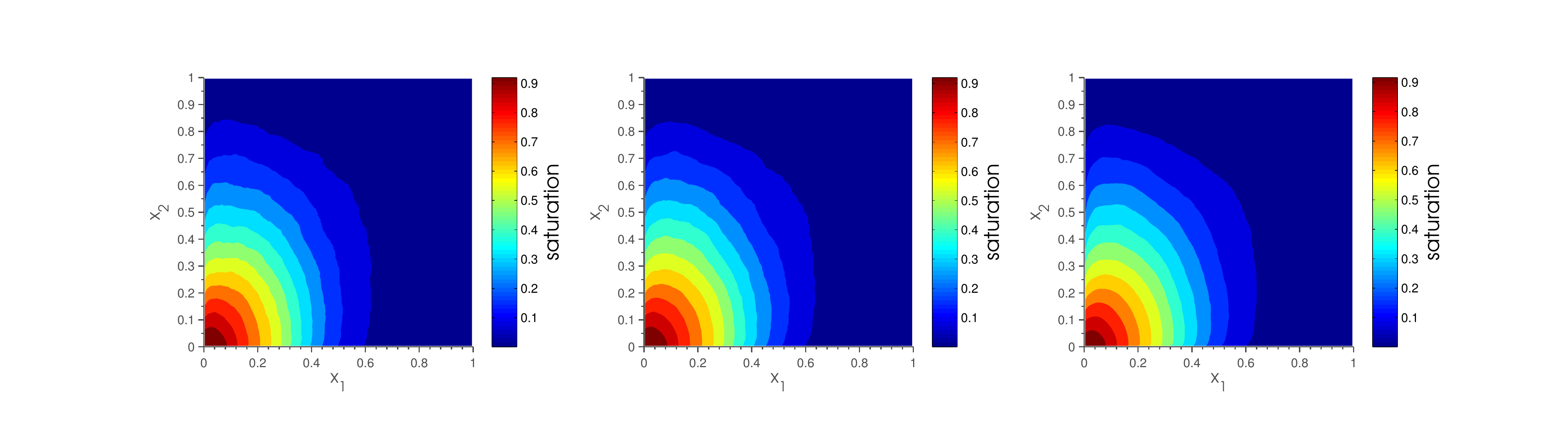}
	         \caption{Comparison between the mean saturation obtained with gFROST (left), FROST (middle) and MCS (right) at time $t = 0.75T$. The stochastic permeability field is highly anisotropic in the $x_2$ direction. The viscosity ratio is set to be $m = 0.25$. $N_s^{MCS} = 3000$ realizations were used for MCS,  $N_s = 1000$ for FROST and $N_s^{g} = 300$ for gFROST.}
	                \label{fig:meanSwComparet3}
	\end{figure*}

\begin{figure*}
		\centering
		\includegraphics[width=\textwidth]{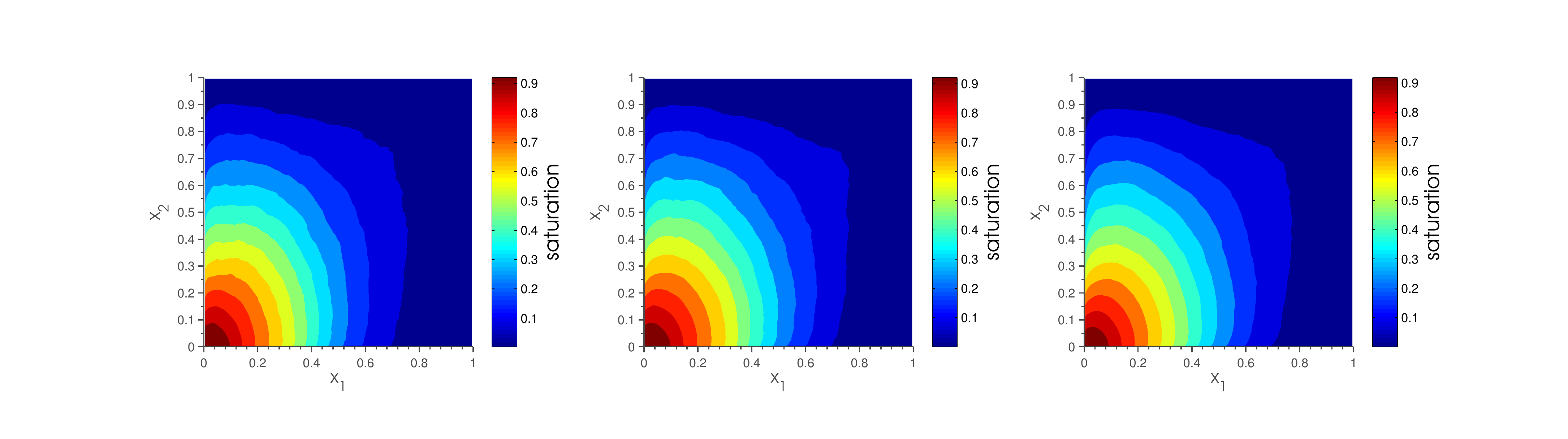}
	         \caption{Comparison between the mean saturation obtained with gFROST (left), FROST (middle) and MCS (right) at time $t = T$. The stochastic permeability field is highly anisotropic in the $x_2$ direction. The viscosity ratio is set to be $m = 0.25$. $N_s^{MCS} = 3000$ realizations were used for MCS,  $N_s = 1000$ for FROST and $N_s^{g} = 300$ for gFROST.}
	                \label{fig:meanSwComparet4}
	\end{figure*}

			Numerically, we compute the integrals by using a simple trapezoidal rule with 400 quadrature points. Results for the mean are displayed in  \autoref{fig:meanSwComparet1}, \autoref{fig:meanSwComparet2}, \autoref{fig:meanSwComparet3}, and \autoref{fig:meanSwComparet4} respectively at times $t = 0.25T$, $t = 0.5T$, $t = 0.75T$, and $t=T$. For each figure, we compare FROST-based, gFROST-based and MCS-based mean saturations. While MCS uses $N_s^{MCS} = 3000$ realizations, FROST uses $N_s = 1000$ realizations and gFROST only uses $N_s^{g} = 300$. For both FROST and gFROST, we use $M_s = 100$ two-time steps solves of transport to estimate $\langle EIT \rangle$. We can see that for all times FROST and MCS are almost undistinguishable, and slight discrepancies with gFROST are noticeable, especially close to the producer. 
		
\begin{figure*}
		\centering
		\includegraphics[width=\textwidth]{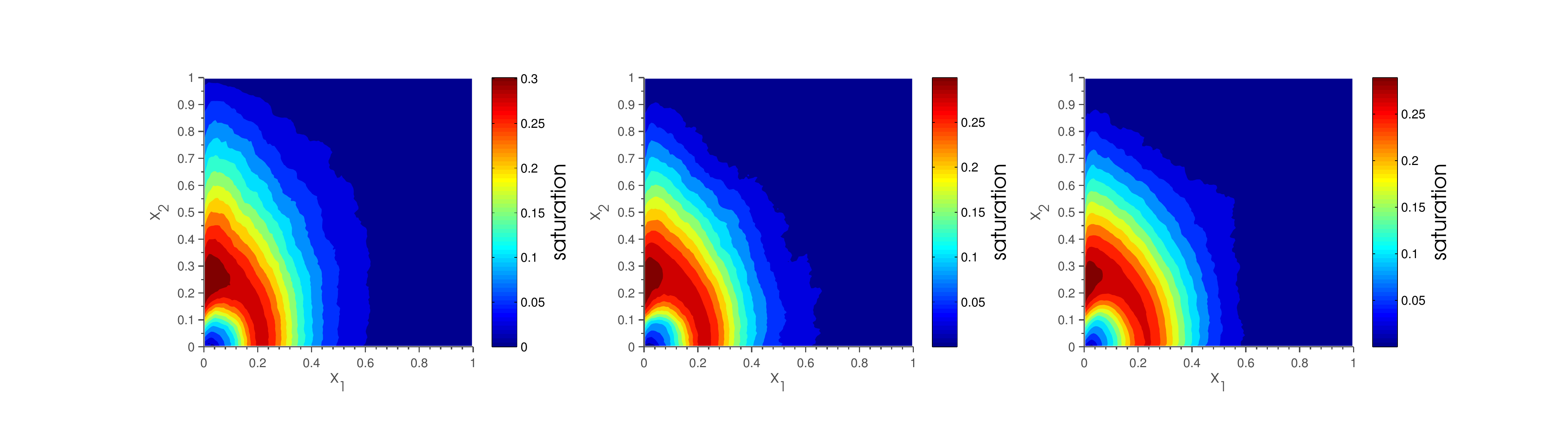}
	         \caption{Comparison between the saturation standard deviation obtained with gFROST (left), FROST (middle) and MCS (right) at time $t = 0.25T$. The stochastic permeability field is highly anisotropic in the $x_2$ direction. The viscosity ratio is set to be $m = 0.25$. $N_s^{MCS} = 3000$ realizations were used for MCS,  $N_s = 1000$ for FROST and $N_s^{g} = 300$ for gFROST.}
	                \label{fig:stdSwComparet1}
	\end{figure*}

\begin{figure*}
		\centering
		\includegraphics[width=\textwidth]{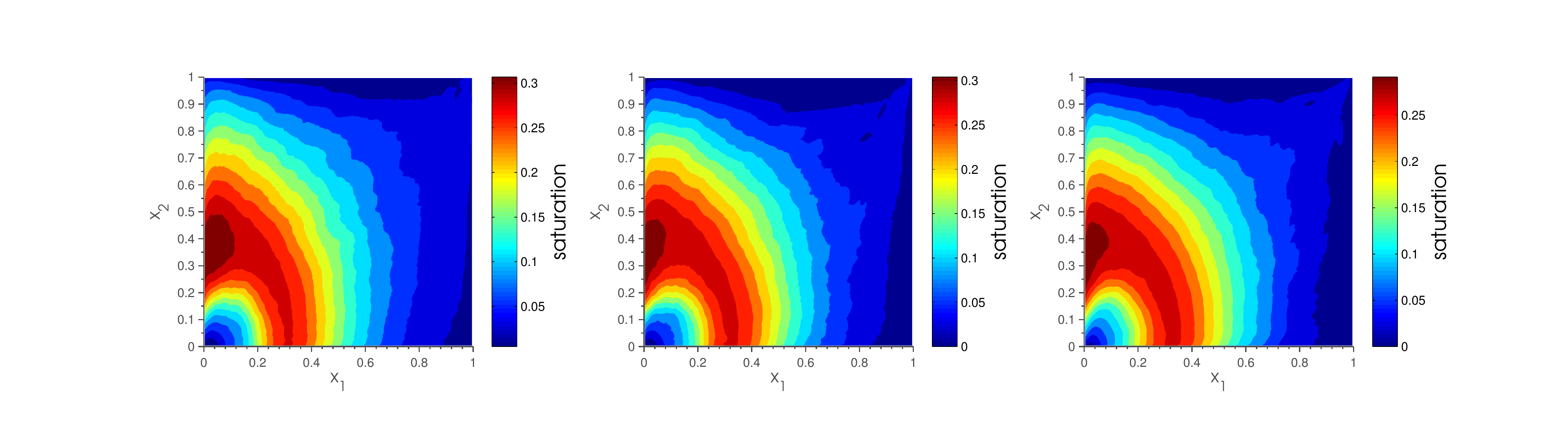}
	         \caption{Comparison between the saturation standard deviation obtained with gFROST (left), FROST (middle) and MCS (right) at time $t = 0.5T$. The stochastic permeability field is highly anisotropic in the $x_2$ direction. The viscosity ratio is set to be $m = 0.25$. $N_s^{MCS} = 3000$ realizations were used for MCS,  $N_s = 1000$ for FROST and $N_s^{g} = 300$ for gFROST.}
	                \label{fig:stdSwComparet2}
	\end{figure*}

\begin{figure*}
		\centering
		\includegraphics[width=\textwidth]{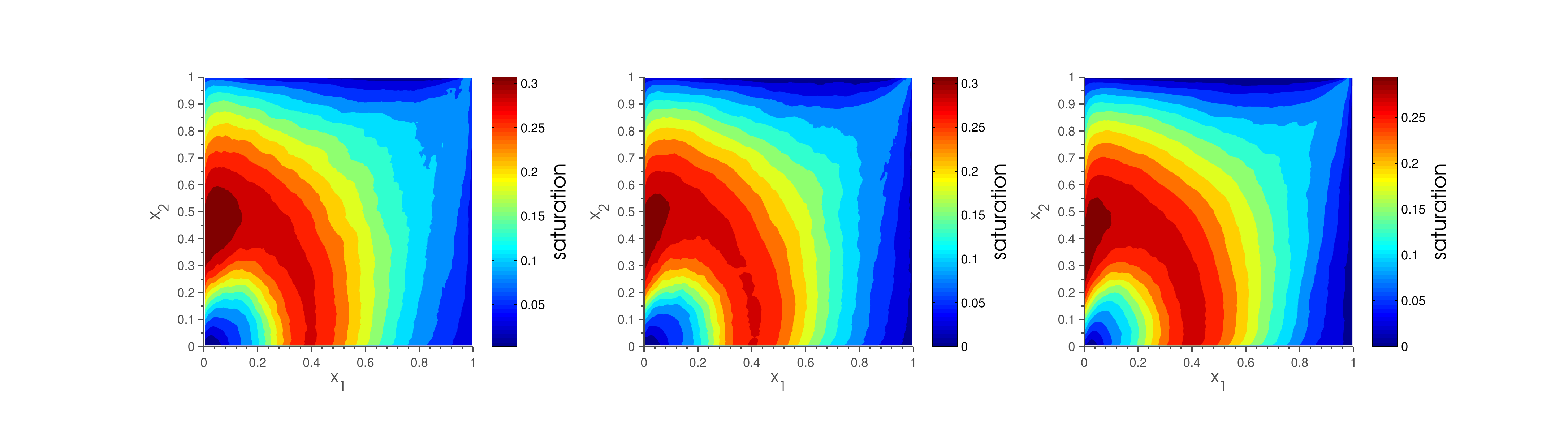}
	         \caption{Comparison between the saturation standard deviation obtained with gFROST (left), FROST (middle) and MCS (right) at time $t = 0.75T$. The stochastic permeability field is highly anisotropic in the $x_2$ direction. The viscosity ratio is set to be $m = 0.25$. $N_s^{MCS} = 3000$ realizations were used for MCS,  $N_s = 1000$ for FROST and $N_s^{g} = 300$ for gFROST.}
	                \label{fig:stdSwComparet3}
	\end{figure*}

\begin{figure*}
		\centering
		\includegraphics[width=\textwidth]{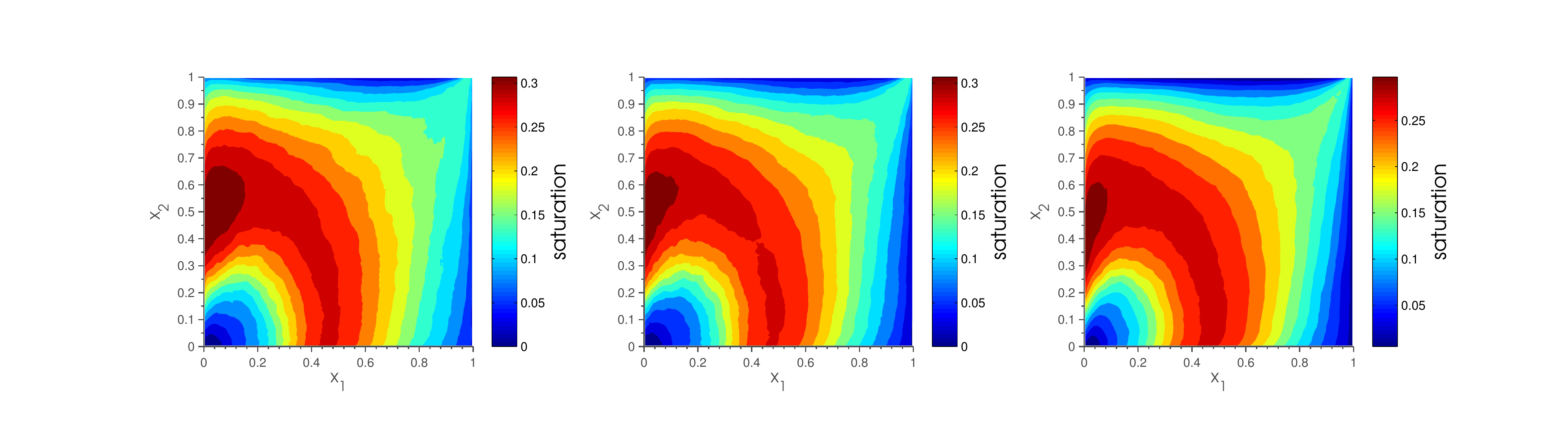}
	         \caption{Comparison between the saturation standard deviation obtained with gFROST (left), FROST (middle) and MCS (right) at time $t = T$. The stochastic permeability field is highly anisotropic in the $x_2$ direction. The viscosity ratio is set to be $m = 0.25$. $N_s^{MCS} = 3000$ realizations were used for MCS,  $N_s = 1000$ for FROST and $N_s^{g} = 300$ for gFROST.}
	                \label{fig:stdSwComparet4}
	\end{figure*}

		Similar illustrations for the standard deviation are depicted in \autoref{fig:stdSwComparet1}, \autoref{fig:stdSwComparet2}, \autoref{fig:stdSwComparet3} and \autoref{fig:stdSwComparet4}. For all times, the match between the three methods is apparent. Again, most discrepancies lie close to the producer. This example showcases the accuracy of the FROST for complex permeability fields, as the saturation standard deviation is non-obvious.

		\subsubsection{Quantiles of saturation}
		
		Mean and standard deviation are useful information for a quick assessment of the average saturation response due to uncertain permeability. However, these statistics do not inform enough about the statistical variability of the saturation, especially because the saturation distribution is generally far from Gaussian. In addition, the non-Gaussianity of the saturation field invalidates the pertinence of a naive confidence interval strategy. Besides, the average saturation field is not suited for decision-making since its diffusive aspect may promote non probable saturation levels. To better quantify the uncertainties, we propose to use a notion that relies more directly on the saturation distribution: saturation quantiles. For $q \in (0,1)$, we define the saturation (first) $1/q$-quantile as follows, 
		\begin{equation}
		\label{eq:quantile}
			[S_w]_q(\vec{x},t) = F_{S_w}^{-1}(q; \vec{x},t).
		\end{equation}
		$[S_w]_q$ corresponds to the likely saturation field such that there is probability $q$ that the true (unknown) saturation field has a smaller saturation at each point $\mathbf{x}$. Or equivalently, with probability $1-q$, the true saturation field is above the saturation $1/q$-quantile. \autoref{eq:quantile} involves solving an inverse problem, which can be done efficiently by any algorithm exploiting the monotonicity of the saturation CDF, $F_{S_w}$, and the one-dimensionality of the problem. 
Nonetheless, in the gFROST case, we can directly evaluate \autoref{eq:quantile} with the help of the inverse of the error function $\Theta$, for which we give a name to emphasize that there are efficient ways to evaluate the inverse directly. 
Indeed, using the Gaussian approximation of log(TOF), within gFROST, the saturation $1/q$-quantile is 
	\begin{equation}
	\label{eq:quantileGSDM}
		[S_w]_q^{}(\vec{x},t) = 
		\check{S}_w\left[ \frac{\exp\left(m^{(N_s^{g})}(\vec{x}) + \sigma^{(N_s^{g})}(\vec{x}) \sqrt{2} \Theta(1 - 2q)\right)}{c t^{\beta}} \right],
	\end{equation}
where $\check{S}_w$ is defined in \autoref{eq:solBL}. 

\begin{figure*}
		\centering
		\includegraphics[width=\textwidth]{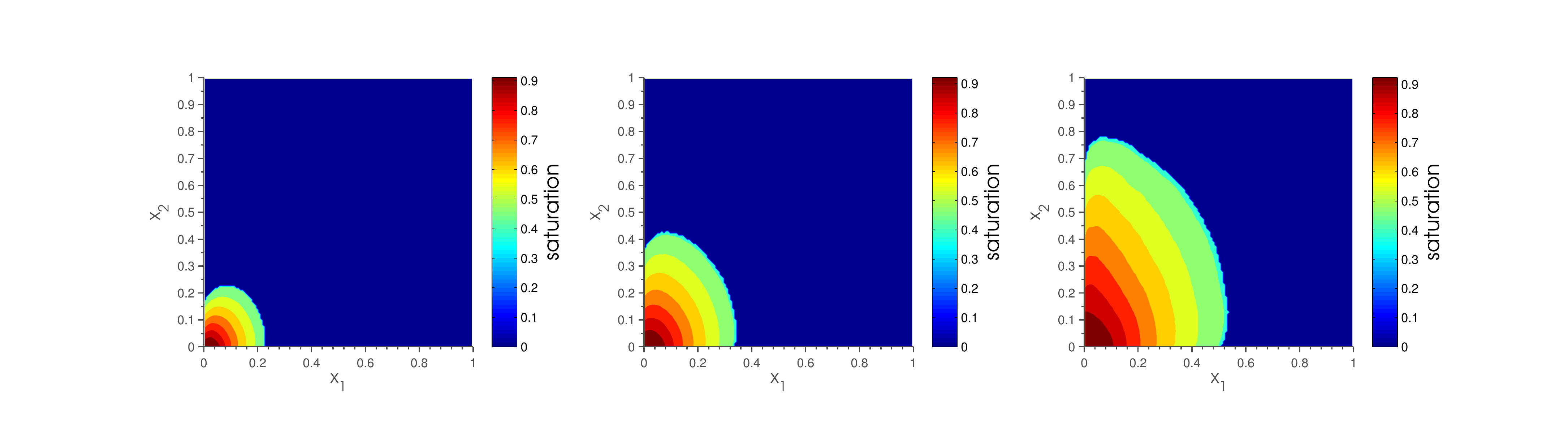}	
		\includegraphics[width=\textwidth]{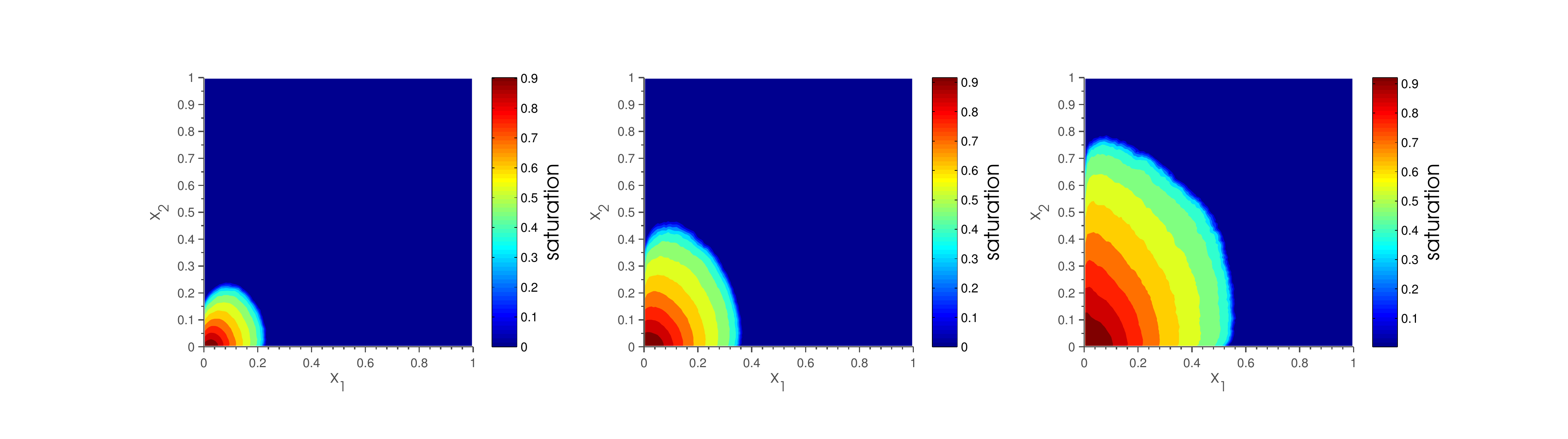}	
	         \caption{Illustration of saturation quantiles with $q = 0.1$ (left), $q = 0.5$ (middle) and $q = 0.9$ (right) at time $t = 0.5T$ with gFROST (top) and MCS (bottom). The stochastic permeability field is highly anisotropic in the $x_2$ direction. The viscosity ratio is set to be $m = 0.25$. $N_s^{MCS} = N_s^g = 3000$ realizations of the $K$ were used.}
	                \label{fig:quantSwComparet1}
	\end{figure*}

\begin{figure*}
		\centering
		\includegraphics[width=\textwidth]{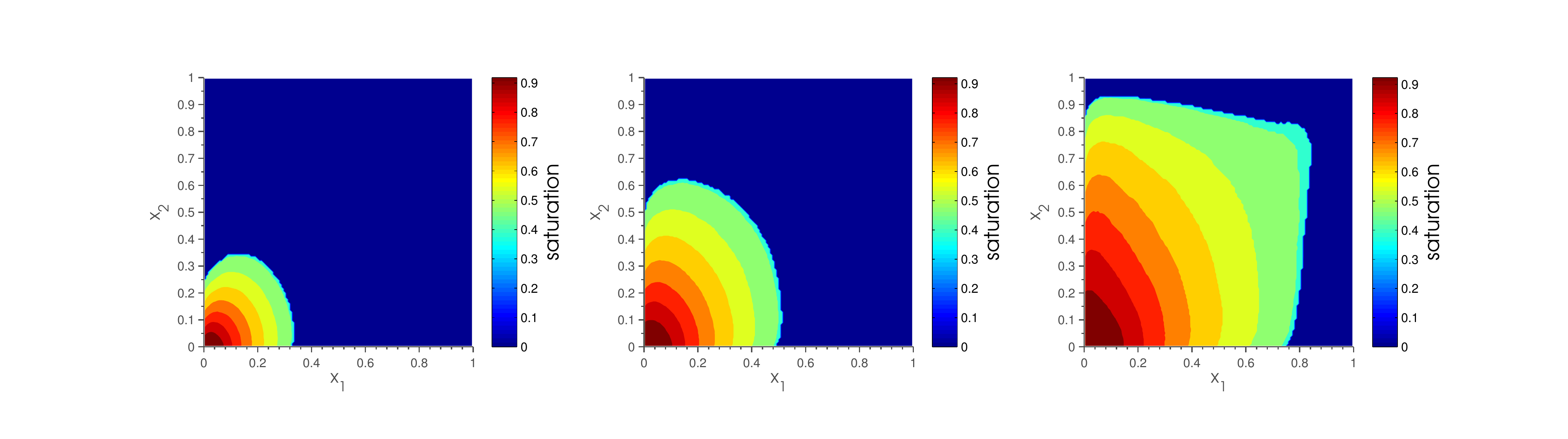}
		\includegraphics[width=\textwidth]{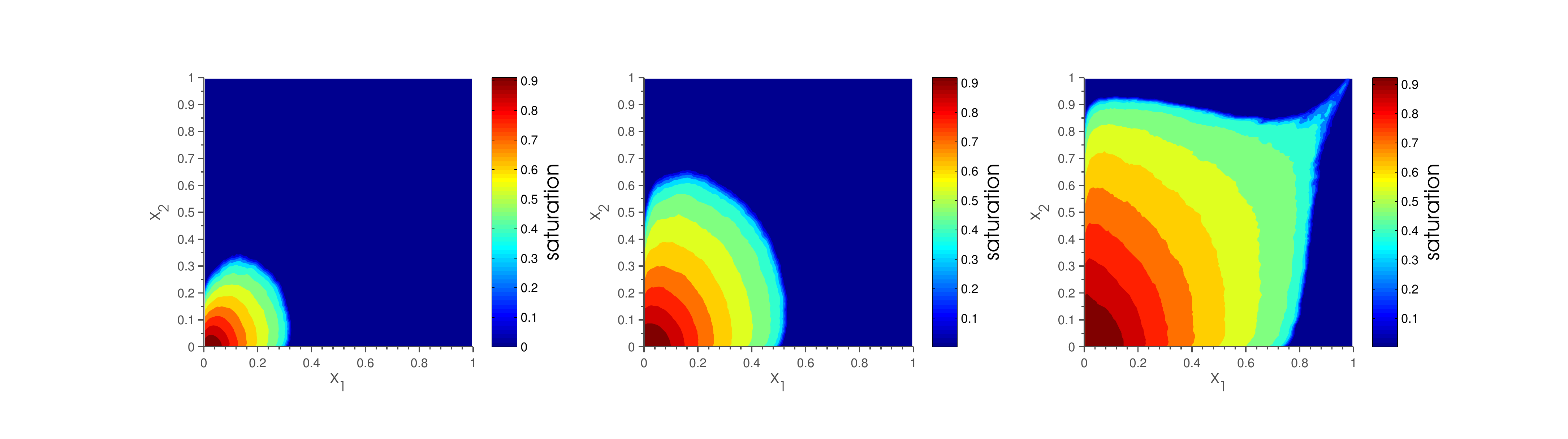}
	         \caption{Illustration of saturation quantiles with $q = 0.1$ (left), $q = 0.5$ (middle) and $q = 0.9$ (right) at time $t = T$ with gFROST (top) and MCS (bottom). The stochastic permeability field is highly anisotropic in the $x_2$ direction. The viscosity ratio is set to be $m = 0.25$. $N_s^{MCS} = N_s^g = 3000$ realizations of the $K$ were used.}
	                \label{fig:quantSwComparet2}
	\end{figure*}

\begin{figure*}
		\centering
		\includegraphics[width=\textwidth]{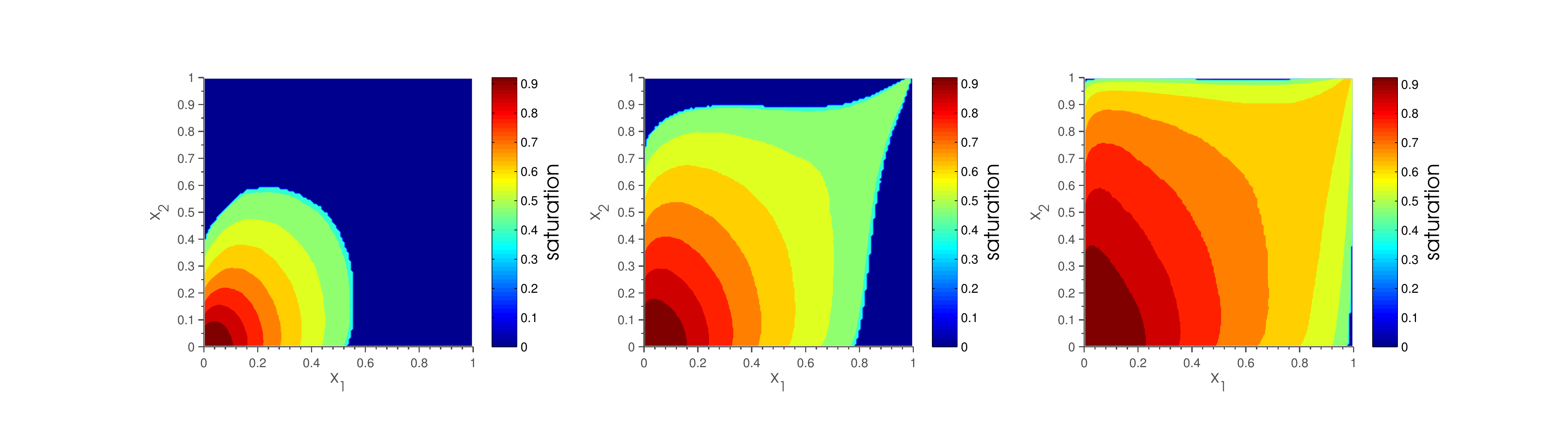}
	         \caption{Illustration of saturation quantiles with $q = 0.1$ (left), $q = 0.5$ (middle) and $q = 0.9$ (right) at time $t = t_{wc}$ with gFROST. The stochastic permeability field is highly anisotropic in the $x_2$ direction. The viscosity ratio is set to be $m = 0.25$. $N_s^{MCS} = N_s^g = 3000$ realizations of the $K$ were used.}
	                \label{fig:quantSwComparet3}
	\end{figure*}

\autoref{fig:quantSwComparet1} and \autoref{fig:quantSwComparet2} compare the likely saturation field for $q = 0.1$, $q = 0.5$ and $q = 0.9$, using the gFROST and MCS, respectively at two different times $t = 0.5T$ and $t = T$. The results are similar and we can see that the gFROST quantiles have sharper fronts than the MCS ones. Informally, if we are concerned about the water breakthrough at the producer, $q = 0.1$ corresponds to the best case scenario, $q = 0.5$ to the median case and $q = 0.9$ to the worst case. For instance, using \autoref{eq:quantileGSDM}, we can estimate the median water breakthrough time $t=t_{wc}$, as it is given by the time the shock front reaches the producer for $q=0.5$, which leads to
\begin{displaymath}
	t_{wc} = \left[ \dfrac{ \exp \left( m^{(N_s^{g})} (\vec{x}_{prod}) \right) }{c \alpha^*} \right]^{1/\beta},
\end{displaymath}
 where $\vec{x}_{prod}$ is the location of the producer. The saturation quantile profiles at time $t=t_{wc}$ are displayed in  \autoref{fig:quantSwComparet3}. This information can be used in risk assessment for instance.

\section{Discussion}
\label{section : disc}


\autoref{eq:multiDCDFm1} provides a simple but reliable estimate of the saturation distribution under stochastic permeability and porosity fields. The main advantage of the proposed FROST method resides in the observation that, in the highly heterogeneous case, the streamlines are essentially fixed. Hence, we can solve the nonlinear aspect of the distribution semi-analytically with a mapping, while the linear part contains the $\log$(TOF) and EIT random fields to be estimated. Moreover, because of its smoothness, the $\log$(TOF) distribution can be efficiently and cheaply estimated by KDE. Therefore, no (or little for the EIT) transport sampling is needed. Besides, for each spatial location, the log(TOF) density is close to Gaussian. This motivates the introduction of an even more efficient sampling method, the gFROST. The gFROST drastically reduces the cost of the distribution method by approximating the distribution of the $\log$(TOF) at each point by a Gaussian distribution with mean and variance given by the mean and variance of the $\log$(TOF). Due to the smoothness of the distribution of $\log$(TOF), only a few samples of the saturation are needed to accurately estimate its mean and standard deviation and consequently form the Gaussian surrogate distribution. 

Within the gFROST, most key statistical quantities are available in a semi-analytical form, such as the saturation quantiles. But equally important is the fact that the EIT can be replaced by its mean, and the mean EIT does not depend on the spatial location and follows a power law in time. Replacing the EIT by its mean requires the EIT standard deviation to be relatively small, which can reduce the range of viscosity ratios and input variance that produce accurate saturation distribution estimates. However, within this range of viscosity ratios, no (or little for determining the EIT parameters) transport sampling is needed. Indeed, the (g)FROST is reusable for all times only at the cost of function evaluations, provided we precompute the power law fitted mean EIT, since no further sampling is needed to estimate the saturation distribution at different times. On the other hand, in MCS, each time step requires to numerically and accurately solve for the nonlinear coupled system comprised of \autoref{eq:incompressible} and \autoref{eq:consmass}. 

Finally, the FROST always produces stable estimates of the saturation PDF because the saturation nonlinearity is taken care of by the analytical mapping and the density estimation is used for the TOF that results from a linear Darcy flow problem and is hence smooth. Direct estimation of the saturation PDF by MCS and KDE, though, is not reliable since the strong nonlinearity of the transport problem results in non-smooth saturation PDFs that KDE cannot manage.

Regarding computational times, on a personal laptop (Mac OS X, $2.7$ Ghz Intel Core i5, $8$ GB memory 1600 MHz), for the numerical example in \autoref{section : numres}, the total running time for MCS with $3000$ realizations is $76,343$ seconds, while it corresponds to $23,067$ seconds for FROST. In other words, we obtain a $3.3$x speed-up mainly due to the fact that, in the FROST case, we only solve for the TOF at initial time. Using FROST with $1000$ realizations reduces the total running time to $7,754$ seconds and gives a speed-up of $9.8$x compared to MCS. Even more, using gFROST with $300$ realizations requires $2,316$ seconds and corresponds to a $32.9$x speed-up compared to MCS. The typical time to run one saturation realization with $N_t=4$ time steps using MCS is $16.5$ seconds, while for one frozen streamline TOF realization it takes $4.6$ seconds (FROST also requires to run full (but few) MCS simulations for $2$ time steps to estimate the mean EIT). KDE estimations are the second most computationally expensive work, but are marginal compared to the cost of running realizations. The MCS-based saturation distribution estimation takes $N_t$ times more than the $\log$TOF distribution estimation ($82$ seconds). Finally, the FROST requires a last step to estimate the saturation distribution and statistics. This last step takes $19$ seconds to output the CDFs, PDFs, means and standard deviations of the saturation for $N_t = 4$ global time steps. But the main advantage of the FROST lies in the observation that only the last step is required to compute saturation distributions and statistics at further times or different sets of times (for instance, if we want to refine the time step and consider $\Delta t = T/N_t$ with $N_t = 10$, we only need to run the last step, which takes $51$ seconds). In contrast, for MCS, we would have to run a new set of realizations to estimate the saturation at refined or further times.

\section{Conclusion}
\label{section : conclusion}

We have designed an efficient distribution method for the water saturation in two-phase flow problems with highly heterogeneous porous media. This strategy extends the ideas presented in \cite{IMT} and demonstrates how we can build fast algorithms to estimate the saturation distribution under uncertain geology for spatial dimensions larger than $1$. 

The (g)FROST method is motivated by the physics of the two-phase flow problem. The developed algorithm benefits from highly heterogeneous porous systems, where the streamline patterns are only weakly time-dependent, and from slow displacement processes (viscosity ratio close to unity). The saturation distribution is estimated by efficiently determining the distribution of the $\log$(TOF) and approximating the EIT by its statistical mean. The FROST method does not suffer from the limitations of perturbation methods in terms of the magnitude of the variance and correlation lengths of geological features, nor from the curse of dimensionality as it does not rely on any spectral expansion. Besides, the method always gives stable estimates of the saturation PDF and CDF. However, it requires to run full MCS for two time steps to estimate the power law for the mean EIT, and the mean EIT approximation is accurate for a restricted range of viscosity ratios. 

In terms of comparing the saturation distributions and moments, the (g)FROST is in good agreement with MCS and is attractive, mainly thanks to its stability, especially for the PDF estimations. Furthermore, the (g)FROST leads to efficient estimates of saturation quantiles, which are more appropriate quantities for uncertainty assessment and decision making. More importantly, the method does not require a specific structure of the input distribution and complex geostatistical models for the permeability and porosity fields can be used without significantly damaging the quality of the estimates.

The FROST is guided by a streamline interpretation of the flow in the porous media, though no actual streamlines are needed to compute the FROST estimates. However, the model relies on the frozen streamline approximation. A streamline solver was used in this study, but examples of the FROST algorithm with a finite volume solver can be found in \citep{IbrahimaTchelepiMeyer2016b}. 

This last remark suggests the first step of future developments. First, the self-similar and analytical aspects of the nonlinear mapping rely on the fact that the initial saturation was considered to be constant in the porous medium. From \autoref{eq:multiDCDFsaturation}, we can see that additional sources of uncertainty can be included in the FROST framework, such as stochastic relative permeability with random coefficients independent from the saturation random field, or constant, but random initial saturation field. However, because these extra sources of uncertainties increase the dimension of the stochastic space to sample, no such tests have yet been studied and further investigation is needed for possible efficient sampling strategies for these cases. For more general initial saturation profiles, we need to rethink the mapping in \autoref{eq:solBL}.

Second, the frozen streamline approximation is valid here for highly heterogeneous media because of the absence of gravity. In problems with gravity, the streamlines are expected to be highly time dependent. In this case, or generally in cases where the frozen streamline approximation is no longer applicable, we could relax this approximation by designing a two step method in which the FROST estimate can be recursively used for a time span and then corrected by re-sampling the TOF, provided that we design an efficient strategy for the first problem, that is the nonlinear mapping with general initial saturation.

Third, the EIT approximation by its mean implies a restriction in fluid viscosity ratios. We can investigate strategies for efficient sampling of $Z$ defined in \autoref{eq:approxZ} instead, thus not requiring any further approximation on the EIT. In parallel, we will investigate the limitations of the method vis a vis the complexity of the geological input distributions. \cite{MC} argue that the permeability field should rather be modeled using training images generated with multiple-point geostatistics, which for instance conserve the geological consistency of channelized systems under uncertainty. We shall therefore investigate the quality of the FROST for these complex, yet essential, input models.

Finally, in terms of algorithm, further considerations are necessary. We shall investigate methods to further reduce or avoid the use of sampling. A first step can be to use stochastic spectral methods to estimate the $\log$(TOF) field. A more involved strategy would consist in building a parametric model for the distribution of the random $\log$(TOF) on the entire domain. All in all, the distribution method presented in this paper is an encouraging step towards designing efficient uncertain quantification tools for large scale flow transport problems and may also be extended to a larger class of physical problems.

\section*{Acknowledgements}

Daniel Meyer is very thankful to Florian M\"uller who kindly provided the streamline code that was applied in this work. Fayadhoi Ibrahima is thankful to Per Pettersson for his constant support in the early stage of the manuscript and to Matthias Cremon for providing SGEMS simulations. The authors are also grateful to the SUPRI-B research group at Stanford University for their financial support.

\bibliographystyle{model2-names} 
\bibliography{Biblio2dDMFay}

\end{document}